\documentclass[12pt,a4paper,twoside]{article}

\usepackage[dvips]{graphics}
\usepackage{epsfig}
\usepackage{cite}

\renewcommand{\d}               {{\rm d}}

\newcommand{\qqbar}  {\ensuremath{\mathrm{q\overline{q}}}}

\newcommand{\epem}   {\ensuremath{e^+e^-}}

\newcommand{\as}     {\ensuremath{\alpha_s}}

\newcommand{\asmu}   {\ensuremath{\alpha_s(\mu)}}

\newcommand{\oaa}    {\ensuremath{\rm{O}(\alpha_s^2)}}

\newcommand{\znull}  {\ensuremath{\mathrm{Z}}}

\newcommand{\rs}     {\ensuremath{\sqrt{s}}}

\newcommand{\chisq}  {\ensuremath{\chi^2}}
\newcommand{\chisqd} {\ensuremath{\chi^2/\mathrm{d.o.f.}}}

\newcommand{\ycut}   {\ensuremath{y_{\mathrm{cut}}}}

\newcommand{\stat}   {\ensuremath{\mathrm{(stat.)}}}
\newcommand{\syst}   {\ensuremath{\mathrm{(syst.)}}}
\newcommand{\tautau} {\ensuremath{\tau^+\tau^-}}

\newcommand{\beq}    {\begin{equation}}
\newcommand{\eeq}    {\end{equation}}
\newcommand{\beeq}   {\begin{eqnarray}}
\newcommand{\eeeq}   {\end{eqnarray}}
\newcommand{\nn}     {\nonumber}

\newcommand{\Tr}     {{\rm Tr}}
\newcommand{\BZ}     {\ensuremath{\cos \chi_{\rm BZ}}}
\newcommand{\NR}     {\ensuremath{\cos \Theta_{\rm NR}}}
\newcommand{\KSW}    {\ensuremath{\cos \Phi_{\rm KSW}}}
\newcommand{\amin}   {\ensuremath{\cos \alpha_{\rm 34}}}
\newcommand{\thrust} {\ensuremath{{\rm thrust}}}

\newcounter{hours}
\newcounter{minutes}

\newcommand{\Printtime}{%
  \setcounter{hours}{\time/60}%
  \setcounter{minutes}{\time-\value{hours}*60}%
  \ifthenelse{\value{hours}<10}{0}{}\thehours:%
  \ifthenelse{\value{minutes}<10}{0}{}\theminutes}

\begin{document}

%
\begin{titlepage}
\begin{center}{\large   EUROPEAN ORGANIZATION FOR NUCLEAR RESEARCH}\end{center}\bigskip
\begin{flushright}
       CERN-EP-2001-001   \\ January 11, 2001
\end{flushright}
\bigskip\bigskip\bigskip

\begin{center}
{\Large\bf  A Simultaneous Measurement of the QCD Colour Factors and the Strong Coupling}\\
\bigskip
\end{center}
\bigskip\bigskip
\begin{center}
{\large The OPAL Collaboration}\\
\bigskip
\end{center}
\bigskip

\begin{abstract}
Using data from \epem\ annihilation into hadrons, taken with the OPAL
detector at LEP at the $Z$ pole between 1991 and 1995, we performed a
simultaneous measurement of the
colour factors of the underlying gauge group of the strong interaction,
$C_F$ and $C_A$, and the strong coupling, $\alpha_s$. The measurement was carried out by fitting
next-to-leading order perturbative predictions to measured angular
correlations of 4-jet events together with multi-jet related variables.
Our results,
\[
C_A=3.02\pm  0.25\stat \pm 0.49\syst\:,\quad
C_F=1.34\pm  0.13\stat \pm 0.22\syst\:,
\]
\[
\as(M_Z)=0.120\pm  0.011\stat \pm 0.020\syst\:,
\]
provide a test of perturbative QCD in which the only assumptions are
non-abelian gauge symmetry and standard hadronization models. The
measurements are in agreement with SU(3) expectations for $C_F$ and
$C_A$ and the world average of $\as(M_{Z})$. 
\end{abstract}

 \bigskip\bigskip\bigskip

\begin{center}{\large
(Submitted to Eur. Phys. J. C)
}\end{center}

\end{titlepage}

\begin{center}{\Large        The OPAL Collaboration
}\end{center}\bigskip
\begin{center}{\small
 G.\thinspace Abbiendi$^{  2}$,
C.\thinspace Ainsley$^{  5}$,
P.F.\thinspace {\AA}kesson$^{  3}$,
G.\thinspace Alexander$^{ 22}$,
J.\thinspace Allison$^{ 16}$,
G.\thinspace Anagnostou$^{  1}$,
K.J.\thinspace Anderson$^{  9}$,
S.\thinspace Arcelli$^{ 17}$,
S.\thinspace Asai$^{ 23}$,
D.\thinspace Axen$^{ 27}$
% ...
}\end{center}\bigskip
\bigskip
$^{  1}$School of Physics and Astronomy, University of Birmingham,
Birmingham B15 2TT, UK
\newline
$^{  2}$Dipartimento di Fisica dell' Universit\`a di Bologna and INFN,
I-40126 Bologna, Italy
\newline
$^{  3}$Physikalisches Institut, Universit\"at Bonn,
D-53115 Bonn, Germany
\newline
% ...
$^{  5}$Cavendish Laboratory, Cambridge CB3 0HE, UK
 \newline
% ...
$^{  9}$Enrico Fermi Institute and Department of Physics,
University of Chicago, Chicago IL 60637, USA
\newline
% ...
$^{ 16}$Department of Physics, Schuster Laboratory, The University,
Manchester M13 9PL, UK
\newline
$^{ 17}$Department of Physics, University of Maryland,
College Park, MD 20742, USA
\newline
% ...
$^{ 22}$Department of Physics and Astronomy, Tel Aviv University,
Tel Aviv 69978, Israel
\newline
$^{ 23}$International Centre for Elementary Particle Physics and
Department of Physics, University of Tokyo, Tokyo 113-0033, and
Kobe University, Kobe 657-8501, Japan
\newline
% ...
$^{ 27}$University of British Columbia, Department of Physics,
Vancouver BC V6T 1Z1, Canada
\newline
% ...
Tel Aviv 69978, Israel.
\bigskip\bigskip\bigskip
\par
Acknowledgements:
\par
We particularly wish to thank the SL Division for the efficient operation
of the LEP accelerator at all energies
 and for their continuing close cooperation with
our experimental group.  We thank our colleagues from CEA, DAPNIA/SPP,
CE-Saclay for their efforts over the years on the time-of-flight and trigger
 systems which we continue to use.  In addition to the support staff at our own
institutions we are pleased to acknowledge the  \\
Department of Energy, USA, \\
National Science Foundation, USA, \\
Particle Physics and Astronomy Research Council, UK, \\
Natural Sciences and Engineering Research Council, Canada, \\
Israel Science Foundation, administered by the Israel
Academy of Science and Humanities, \\
Minerva Gesellschaft, \\
Benoziyo Center for High Energy Physics,\\
Japanese Ministry of Education, Science and Culture (the
Monbusho) and a grant under the Monbusho International
Science Research Program,\\
Japanese Society for the Promotion of Science (JSPS),\\
German Israeli Bi-national Science Foundation (GIF), \\
Bundesministerium f\"ur Bildung und Forschung, Germany, \\
National Research Council of Canada, \\
Research Corporation, USA,\\
Hungarian Foundation for Scientific Research, OTKA T-029328,
T-023793 and OTKA F-023259.\\

\newpage

\section{Introduction}
Electron-positron annihilation into hadrons at high energies provides a
precise means to test Quantum Chromodynamics (QCD), the theory of
strong interactions. Due to the purely leptonic initial state, there
are many experimental quantities for which the long-distance
(non-perturbative) effects are modest, especially when compared to similar
quantities in hadron-hadron collisions or deep inelastic scattering. 
These quantities, for instance the total cross section and jet-related
correlations, can be calculated in perturbative QCD as a function of a
single parameter, the strong coupling strength, \as. Therefore, many
QCD tests based on measurements of $\as$ have been carried out at LEP,
the Large Electron-Positron collider at CERN
(see, for example, ~\cite{as_ALEPH,as_DELPHI,as_L3,as_OPAL,as_OPAL_color,as_pp}).  

A key ingredient of QCD is the underlying gauge group. The
simultaneous measurement of the strong coupling and the eigenvalues of
the quadratic Casimir operator of the underlying gauge theory, the
$C_F$ and $C_A$ colour factors, provides a more general and
comprehensive test of QCD than measurements of $\as$ alone. The
possible existence of a light gluino\footnote{The exclusion of gluinos with mass between 3--5\,GeV has been debated in the literature recently~\cite{gluino,cs-f,color_ALEPH2}.}
influences both $\as$ and the measured value of the colour factors (or,
assuming SU(3) dynamics, the number of light fermionic degrees of
freedom $N_f$).
 A simultaneous
fit of these parameters to data provides a means to investigate whether the data
favour or exclude the additional degrees of freedom.

Previous analyses of LEP data have been performed to probe the
underlying  gauge structure of the
theory~\cite{a34,color_ALEPH1,color_DELPHI,color_OPAL,color_ALEPH2}.
 The leading order perturbative prediction was
fitted to distributions of four-jet angular correlations (normalised to
the total number of the selected four-jet events), leaving the ratios
of the colour factors as free parameters. Due to the normalization,
the leading order prediction of these distributions is independent of
the strong coupling. However, the independence of the leading order prediction
from the strong coupling does not imply that the radiative corrections
are negligible. In fact, in Ref.~\cite{NTangulars} the value of $T_R/C_F$ (see the Appendix) was estimated to increase about 25\% if next-to-leading order (NLO) theoretical predictions are used instead of the leading order (LO) ones. It is therefore desirable to 
explicitly verify the effect of the next-to-leading order corrections on these
normalised angular distributions.

Beyond leading order in the theoretical predictions, the
strong coupling and the colour factors are interdependent. On the one hand,
even the normalised next-to-leading order predictions for the angular
distributions depend upon \as~\cite{NTangulars}, while on the other
hand, \as\ itself depends on the colour factors through its running.
In this sense, beyond leading order, the only consistent way to determine
the three free parameters is a simultaneous measurement of the strong
coupling and the colour factors.
 
Predictions beyond leading order are made possible by theoretical
developments achieved in the last few years. For multi-jet rates as
well as numerous event shape distributions with perturbative expansions
starting at O(\as), matched next-to-leading order and next-to-leading
logarithmic approximations provide very precise descriptions of the
data over a wide range of the available kinematic region
\cite{CTWT,CDOTW,DSjets,NTqcd98}. Also, recently, the next-to-leading
order predictions for the distributions of four-jet angular
correlations have been calculated \cite{NTangulars,Signer}. To make use
of these developments, we perform a simultaneous fit of the strong
coupling and ratios of colour factors using next-to-leading order
predictions of four-jet rates, differential two-jet rates and four-jet
angular correlations. The data were collected by the OPAL
Collaboration at LEP.

The outline of the paper is as follows: In Section~2, we present the
observables used in the analysis and describe their best available
perturbative predictions. The analysis procedure is explained in detail
and the fit results are presented in Section~3. Section~4 contains the
discussion of the systematic checks which were performed and the corresponding systematic errors. We collect our results
and compare them to the results of previous studies in Section~5. Section~6
contains our conclusions.

\section{Observables}

To perform a simultaneous measurement of the strong coupling and the
colour factors
we use multi-jet related variables to gain sensitivity to
\as\, and four-jet angular correlations to gain sensitivity to the colour
factors.

There are many ways of defining jets. For the present analysis we use
the Durham scheme~\cite{CDOTW,Catani}. Starting by defining each
particle to be an individual jet, a resolution variable $y_{ij}$ is
calculated for each pair of jets $i$ and $j$:
\beq
y_{ij}=\frac{2\,\min(E_i^2,E_j^2)\,(1-\cos \theta_{ij})}{E_{\rm vis}^2}\:,
\eeq
where $E_i$ and $E_j$ are the energies of jets $i$
and $j$, $\theta_{ij}$ is the angle between them and $E_{\rm vis}$ is
the sum of the energies of the visible particles in the event or of the partons in a theoretical calculation.
If the smallest
value of $y_{ij}$ is smaller then a predefined value $\ycut$, the pair
is replaced by a pseudo-jet with four momentum
$p_{ij}^\mu =  p_i^\mu + p_j^\mu$ and the clustering starts again with
momenta $p_i^\mu$ and $p_j^\mu$ dropped and $p_{ij}^\mu$ added to the
final state. Clustering ends when the smallest value of $y_{ij}$ is
larger than $\ycut$.

\begin{figure}[!htb]
\begin{center}
\vspace*{-0.3cm}
\epsfig{file=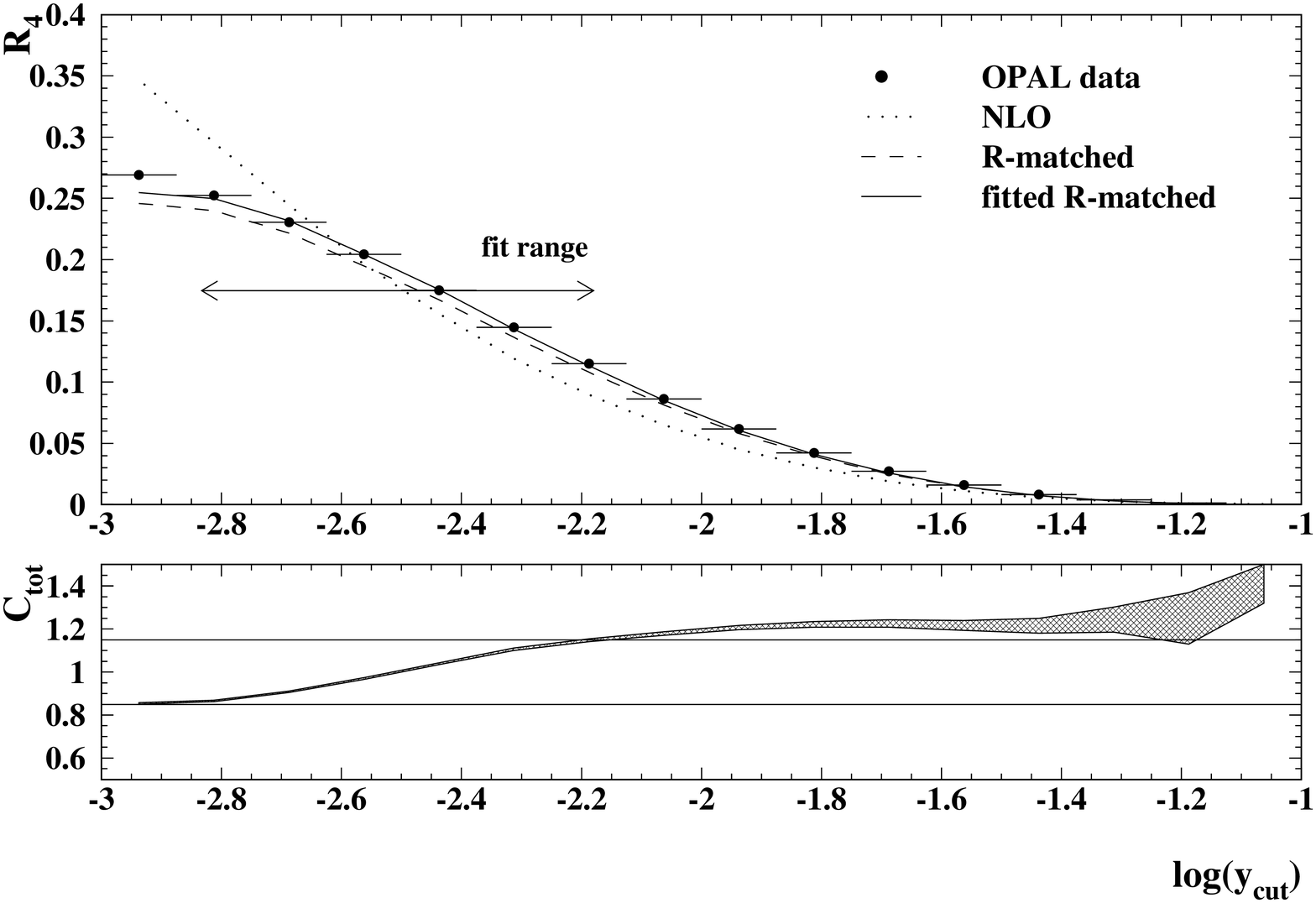,clip,width=140mm}\\
\vspace*{-0.5cm}
\epsfig{file=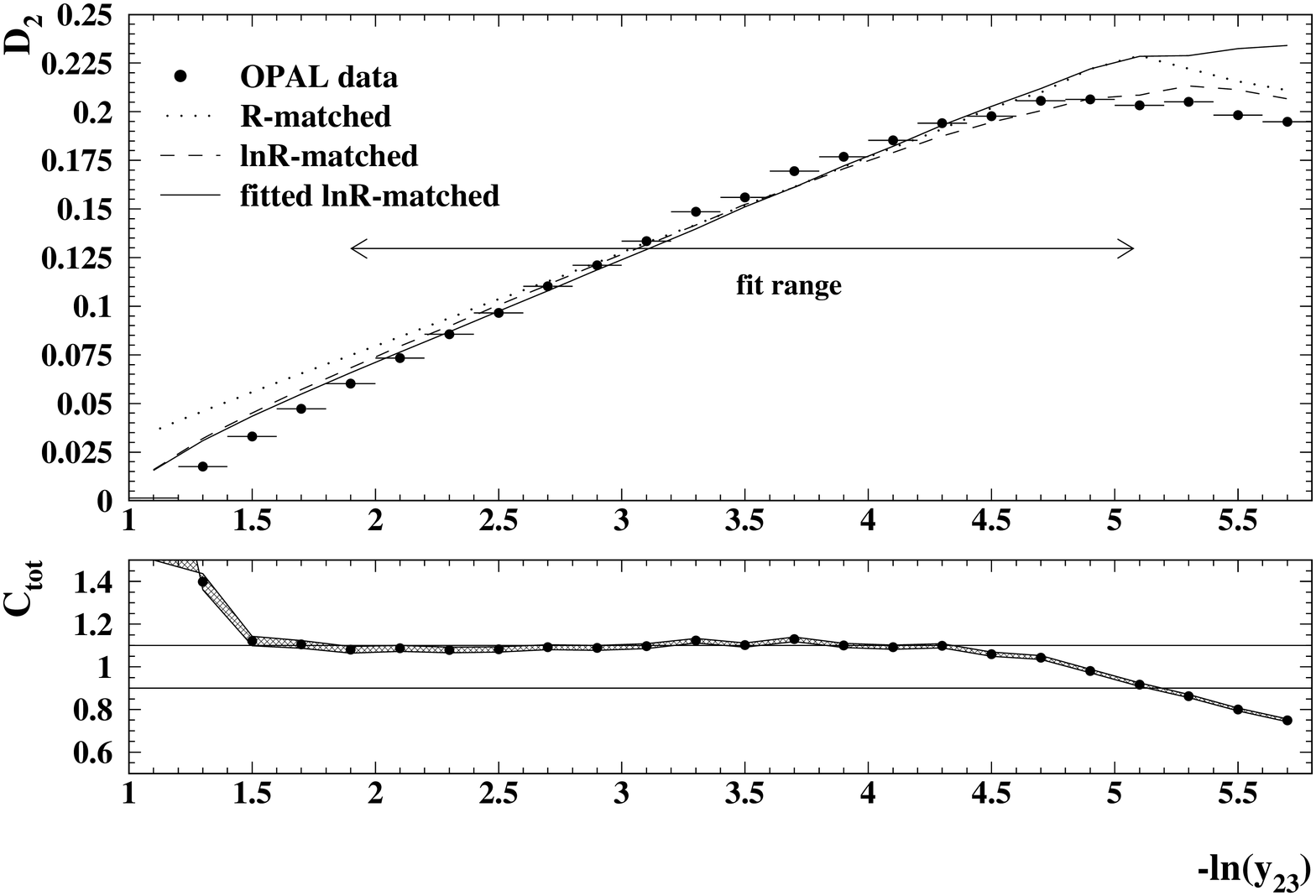,clip,width=140mm}
\vspace*{-0.3cm}
\caption[fit]{Distributions of the four-jet rates and the differential
two-jet rates. OPAL data corrected to the parton level are denoted by points. The
statistical uncertainty in the bins is smaller than the size of the
circles.  The small sections below each plot indicate the correction
factors $C_i^{\rm tot}$ for experimental and hadronization effects with their statistical error (shaded area). The horizontal lines with arrows indicate the selected fit region.
The QCD predictions shown by the dotted and dashed lines employ
the
world average $\as=$0.119 and the standard values $C_A=3$ and $C_F=4/3$.
The solid lines represent the result of the simultaneous fit of all six variables.
}
\label{f:rates}
\end{center}
\end{figure}

In our analysis we use the differential two-jet rate,
\beq
D_2(y_{23})\equiv
\frac{1}{\sigma_{\rm tot}}\,\frac{\d \sigma}{\d y_{23}}\:,
\eeq 
where $y_{23}$ is the \ycut\ value for which the two- and
three-jet configurations are separated in a given event, and the four-jet rate,
\beq
R_4(\ycut)\equiv
\frac{\sigma_{4-{\rm jet}}(\ycut)}{\sigma_{\rm tot}}\:,
\eeq
to gain sensitivity to \as. For theory, $\sigma_{\rm tot}$ is the total
hadronic cross section at O(\as) accuracy. For experiment, $\sigma_{\rm
tot}$ is the total visible hadronic cross section. The $D_2$
distribution has been extensively used at LEP to determine
\as~\cite{as_OPAL,d2_as1}. Use of $R_4$ allows us to obtain
sensitivity to \as\ using a four-jet based variable. For these
observables both next-to-leading order (NLO) \cite{DSjets,NTqcd98} and
next-to-leading logarithmic (NLL) ~\cite{CDOTW}
perturbative predictions are known. The former provide a fixed order
approximation valid for large values of the resolution variable, while
the latter provide an all order summed but approximate cross section
valid for small values of the resolution variable. The explicit NLO
and NLL formulae  and their matched expressions used in our fits are
given in the Appendix. Using those formulae, we obtain various theoretical
predictions for the four-jet rates ($R_4$) and the differential two-jet
rates ($D_2$) which are shown in Fig.~\ref{f:rates}. In these plots,
the theoretical predictions with the exception of the fitted curves were
obtained using the world average of the strong coupling $\as(M_Z) =
0.118$ \cite{as} and standard QCD values for the colour factors,
$C_F = 4/3$ and $C_A = 3$. The theoretical predictions in
Fig.~\ref{f:rates} are shown in comparison to OPAL data corrected to
the parton level. The manner in which the data are corrected to the
parton level is explained in Section~\ref{corr}. It is seen that the
theoretical predictions provide a good description of the data. For the theoretical plots in Fig.~\ref{f:rates}, we used the $R$-matched expression for
$R_4$ and $\ln R$-matched expression for $D_2$ (see the Appendix).

The other class of observables we employ in our measurements is the
four-jet angular correlations. The angular variables are 
\begin{itemize}
\item the Bengtsson-Zerwas angle \cite{BZ}:
$\chi _{BZ}
=\angle [(\vec{p}_{1}\times \vec{p}_{2}),(\vec{p}_{3}\times \vec{p}_{4})]$;
\item the modified Nachtmann-Reiter angle \cite{NR}: 
 $\Theta _{NR}=\angle[(\vec{p}_{1}-\vec{p}_{2}),(\vec{p}_{3}-\vec{p}_{4})]$;
\item the K\" orner-Schierholtz-Willrodt angle \cite{KSW}:\\
$\Phi _{KSW}=\frac{1}{2}
\left(\angle[(\vec{p}_{1}\times \vec{p}_{4}),(\vec{p}_{2}\times \vec{p}_{3})]
     +\angle[(\vec{p}_{1}\times \vec{p}_{3}),(\vec{p}_{2}\times \vec{p}_{4})]
\right)$;
\item the angle between the two lowest energy jets \cite{a34}:
$\alpha _{3,4}=\angle [\vec{p}_{3},\vec{p}_{4}]$,
\end{itemize}
where the $\vec{p}_i$ denote the three-momenta of the energy-ordered jets
$(E_1>E_2>E_3>E_4)$. We defined the jets using the Durham clustering
and selected four-jet events at $\ycut=0.008$. In order to have a
high statistics four-jet sample, \ycut\ should be chosen in the
range from $10^{-2}$ to $10^{-3}$. However, we cannot choose \ycut\ to be
much smaller than 0.01 for technical reasons (see Section~3.3.1).
For the four-jet events, used in obtaining the angular correlations,
an additional cut was placed on the energy of each jet: $E>E_{\rm min}=3\,$GeV.

\begin{figure}[!htb]
\begin{center}
\epsfig{file=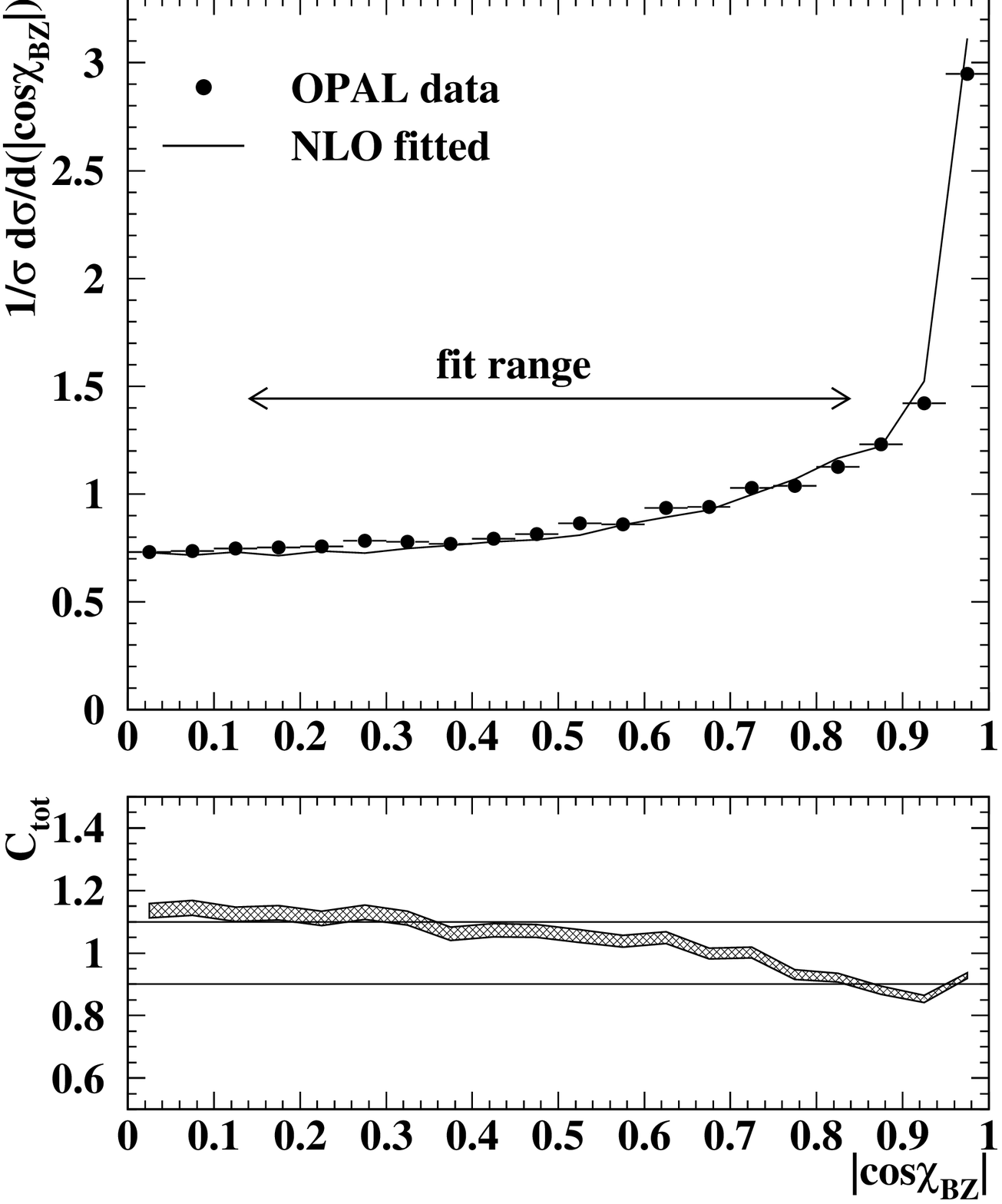,clip,width=75mm}
\epsfig{file=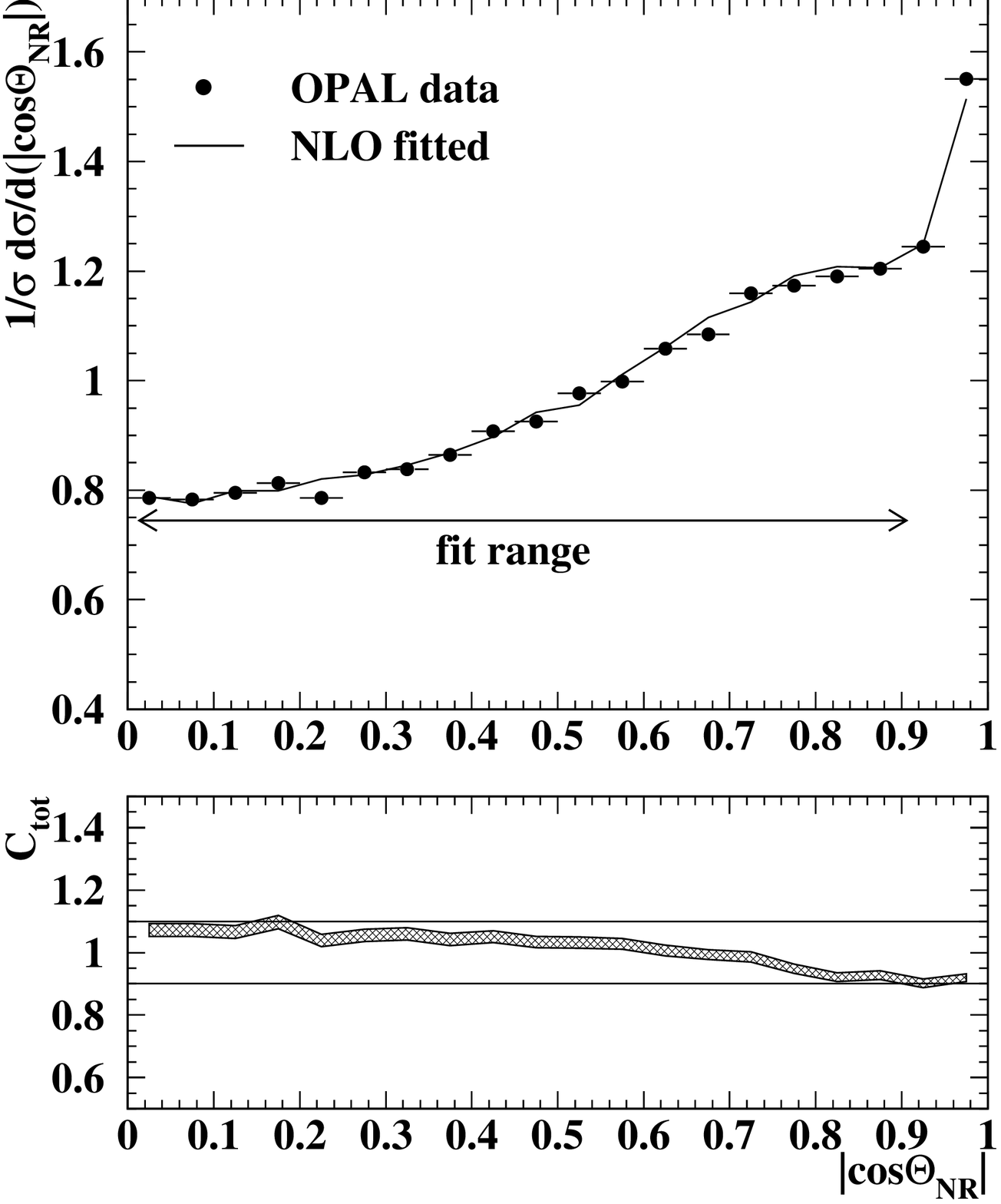,clip,width=75mm}\\
\epsfig{file=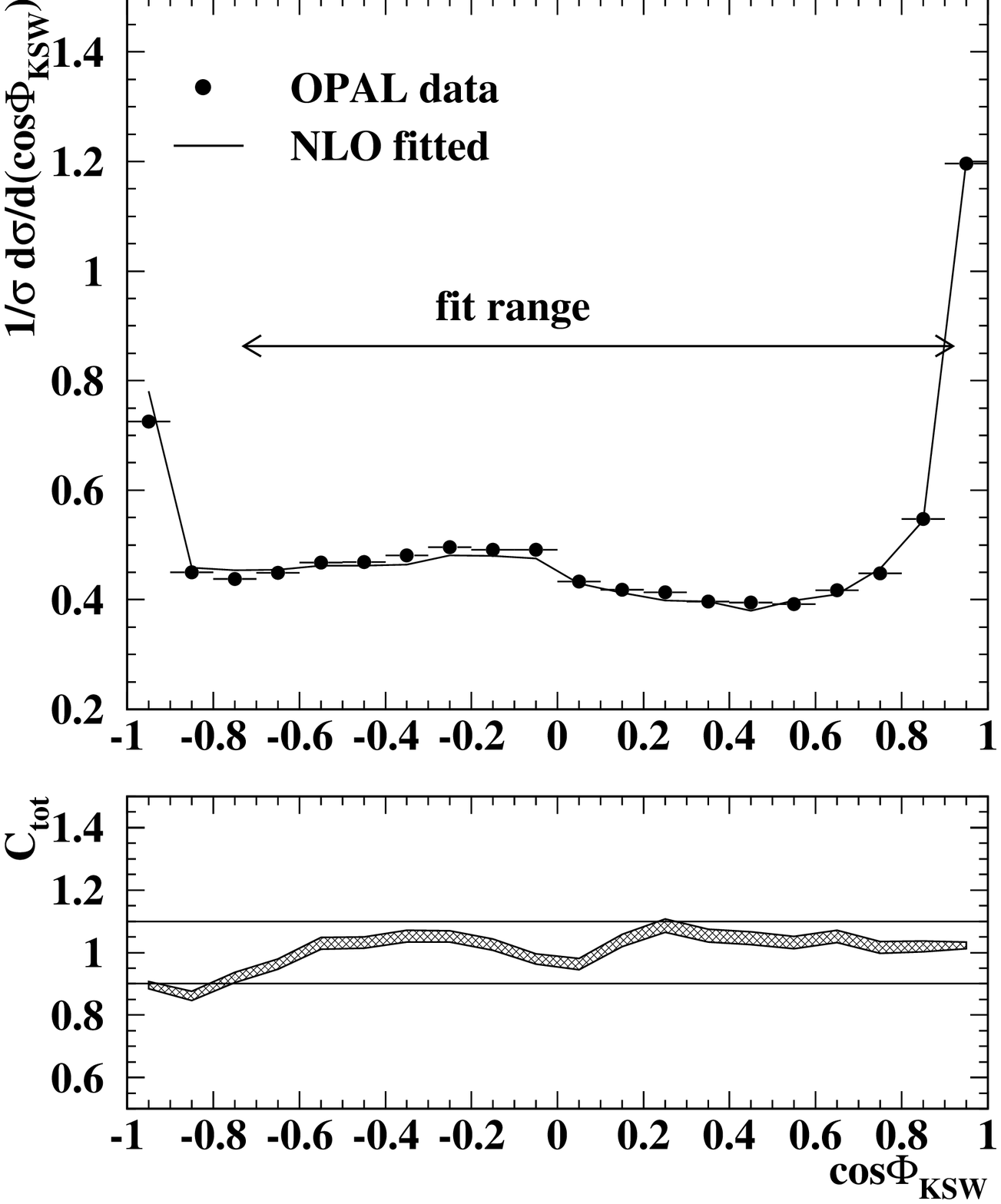,clip,width=75mm}
\epsfig{file=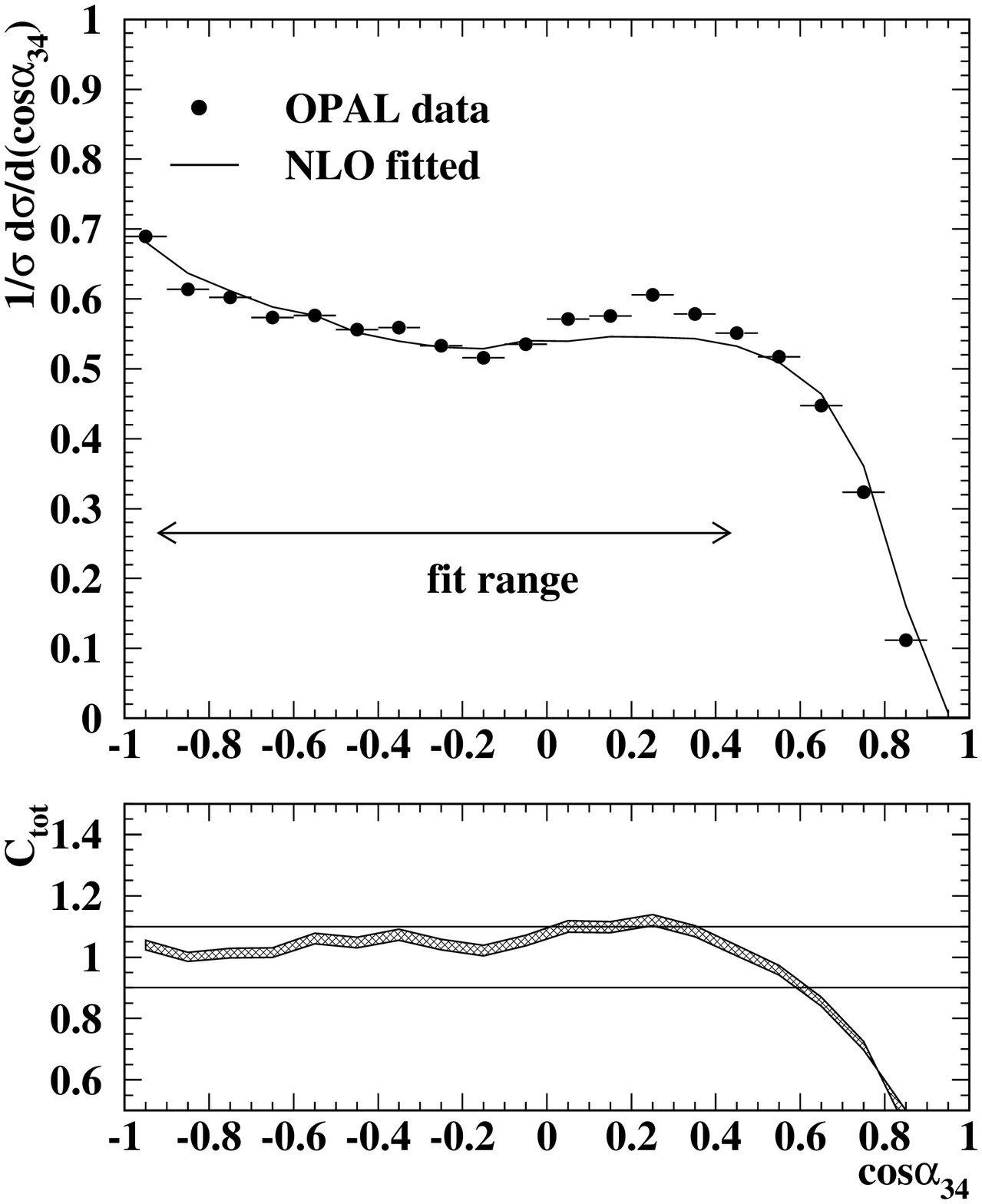,clip,width=75mm}
\caption[fit]{Distributions of the four-jet angular correlations.  OPAL
data corrected to the parton level are denoted by points. The
statistical uncertainty in the bins is smaller than the size of the circles.
The small sections below each plot indicate the correction factors
$C_i^{\rm tot}$ for experimental and hadronization effects with their statistical error (shaded area). The horizontal lines with arrows indicate the selected fit region.
For the normalised angular correlations, the distributions at leading and
next-to-leading order are almost the same \cite{NTangulars,Signer},
therefore, only the NLO results are shown.
}
\label{f:angles}
\end{center}
\end{figure}

To obtain as much information as possible, we used all four angular
variables in our analysis. The shape of the distributions of these
angular correlations at LO and NLO are hardly distinguishable. 
Fig.~\ref{f:angles} shows the normalised NLO theoretical predictions
fitted to the data.

Although the various angular correlations are not entirely independent,
the correlations (see Section~\ref{sec_fit}) are not large as can be seen
from Table~\ref{t:corr}, where the maximum values of the magnitudes of
the bin-wise correlations between the observables are given.  We also
performed fits using a single angular variable together with one of the
jet-related variables and we observed that these fits give very
scattered central values. However, in these cases we found very strong
correlations between the fitted parameters ($\simeq$ 1), indicating
that the distributions do not contain sufficient information to constrain all of the QCD parameters simultaneously.
\begin{table}[!htb]
  \begin{center}
    \begin{tabular}{|c||ccc|} 
\hline
      &  \NR   &  \KSW  &  \amin  \\
\hline\hline
\BZ   &  0.46  &  0.34  &  0.45   \\
\NR   &        &  0.37  &  0.37   \\
\KSW  &        &        &  0.44   \\
\hline
   \end{tabular}
  \end{center}
  \caption[]{
    Maximum values of the bin-wise correlations between the angular
correlation variables. }
\label{t:corr}
\end{table}

\section{ Analysis Procedure and Results}

\subsection{ The OPAL detector}
\label{sec_detector}
A detailed description of the OPAL 
detector can be found in Ref.~\cite{OPALtech}. This analysis
relies mainly on the reconstruction of charged particle trajectories
and on the measurement of energy deposited in the electromagnetic and
hadronic calorimeters.

The central detector contains a silicon micro-vertex detector and three
drift chamber devices: an inner vertex chamber, a large jet chamber and
a surrounding $z$-chamber. The central detector is located inside a
solenoidal magnet which provides a uniform axial\footnote{In the OPAL
coordinate system the $+x$ axis points towards the centre of the LEP
ring, the $y$ axis points approximately upwards and the $z$ axis points
in the direction of the electron beam. The polar angle $\theta$ and the
azimuthal angle $\phi$ are defined with respect to $z$ and $x$,
respectively, while $r$ is the distance from the $z$-axis.} magnetic
field of approximately 0.435 T.  Tracking of the charged particles is
performed with this detector. Most of the tracking information is
obtained from the jet chamber which provides up to 159 measured space
points per track, and close to 100\,\% tracking efficiency in the region
$|\cos \theta|<0.98$. The average angular resolution is approximately
0.1\,mrad in $\phi$ and 1\,mrad in $\theta$.

Electromagnetic energy is measured by lead glass calorimeters
surrounding the magnet coil, separated into a barrel ($|\cos
\theta|<0.82$) and two end-cap ($0.81<|\cos \theta|<0.98$) sections. The
electromagnetic calorimeter consists of 11\,704 lead glass blocks with a
depth of 24.6 radiation lengths in the barrel and more than 22
radiation lengths in the end-caps. This calorimeter is surrounded by
the hadronic calorimeter of the sampling type measuring the energy of
hadrons emerging from the electromagnetic calorimeter.

\subsection{ Data samples }
\label{sec_datamc}

The analysis presented in this paper is based on 4.1 million 
hadronic events recorded within 3 GeV of the \znull\ peak by the OPAL detector between 1991 and 1995.
The descriptions of the OPAL trigger system and the offline
multihadronic event selection are given in Refs.~\cite{OPALtrig} and
\cite{OPALsel}. In this analysis, additional criteria were applied to
eliminate poorly measured tracks and obtain well contained events.

Each track was required to have (i) at least 40 measured
points in the jet chamber, (ii) transverse momentum in the $r$--$\phi$ plane
greater than 0.15\,GeV/$c$, (iii) measured momentum less than 60\,GeV/$c$,
(iv) $|\cos \theta|<0.97$, (v) distance of closest approach to the
origin in the $r$--$\phi$ plane of no more than 5\,cm, (vi) and less than
25\,cm in the $z$ direction.  Hadronic events were required to contain at
least five tracks to reduce contamination from {$\epem\rightarrow\tautau$} events.

Clusters in the electromagnetic calorimeter were accepted if they had more than 0.1\,GeV
energy in the barrel section or more than 0.25\,GeV in the end-cap
section. The corresponding clusters were required to span
at least two lead glass blocks. Clusters in the hadronic calorimeter
were required to have energies larger than 1\,GeV. 

For the calculation of the visible energy, $E_{\rm vis}$, the tracks are assumed to have the pion mass, and clusters are treated as photons.
To determine the energy of each cluster measured in the electromagnetic
and hadronic calorimeters, we used a matching algorithm~\cite{mt}. This
algorithm reduces double counting of energy and gives a better
resolution both in energy and angle than the calorimeters alone. We
did not impose energy-momentum conservation, but checked that such a
constraint does not affect our results.

The event thrust axis was determined using the accepted tracks
and clusters.  In order that the event be well contained, we required
$|\cos \theta_{\thrust}|<0.9$. 

The final event sample contained about 3.6 million hadronic \znull\ decays from which about 250\,000 were identified as four-jet events.

\subsection{Data Correction}
\label{corr}
We choose as the theoretical reference the distributions given in Eqs.~(\ref{difflogR}), (\ref{Rmatch}) of the Appendix
which are obtained from a
parton level calculation. In order to compare our (detector level)
data to these (parton level) theoretical predictions the measured
distributions were corrected for the effects of the detector and
hadronization.  

First, we combined the distributions of the six variables $\BZ,\NR, \KSW,
\linebreak 
\amin, D_2$ and $R_4$ into a
one-dimensional distribution,
with 128 bins, 20 for each angular correlation and 24 each
for the four-jet rates (covering the range of $0.001<y_{cut}<1.0$) and
the differential two-jet rates (in the range of $1.0<-\ln y_{23}<5.8$).
Next, we applied bin-by-bin corrections for detector distortions and
hadronization, as described below. The correction factors were obtained
from the same distributions computed from Monte Carlo events.

\subsubsection{Monte Carlo simulation}

The correction factors were estimated using two different modes of the JETSET Monte
Carlo program~\cite{jetset,jetset3}: the parton shower mode and the matrix element mode. In the parton shower approach, the evolution of the parton system is treated as a branching process based on the leading logarithmic approximation starting from the original \qqbar\ pair. In the matrix element mode, up to four partons are generated according to the second-order ERT calculation~\cite{ert}.

It is well known that the parton shower versions of the available Monte
Carlo programs provide a good description of jet rates, but cannot
describe the structure of the four-jet events properly~\cite{Cowan}. Therefore, to determine correction factors for the
four-jet angular correlations, we used Monte Carlo events with only
four partons (\qqbar gg and \qqbar\qqbar) in the final state.  These
events were generated using the second-order matrix element (ME)
calculation, subjected to hadronization, as implemented in JETSET
version 7.4 assuming standard QCD colour
factors and the set of parameters ERT-MC-1 of Ref.~\cite{param}. 
560\,000 Monte Carlo events were generated.  To avoid singularities in
the four-parton generation of the LO matrix elements, an intrinsic
cutoff $y_0$ has to be applied in the Monte Carlo simulation. In
JETSET, $y_0=0.01$ is used, calculated according to the invariant mass
resolution definition, which for four-jet events corresponds to a Durham
resolution of about 0.004 (yielding the same four-jet rates).  Our
resolution criterion of $\ycut = 0.008$ is safely larger than the
intrinsic cut-off.  For normalised angular correlations, the
distributions at leading and next-to-leading order are almost the same
\cite{NTangulars,Signer}.  Thus for the purpose of determining the
corrections the ERT Monte Carlo events are sufficient.

For the correction  of the differential two-jet rates and the four-jet
rate, we used 2 million events generated with the parton shower (PS)
version of JETSET, using parameters tuned to OPAL data at $\sqrt{s}=$91
GeV described in Ref.~\cite{psparam}. For both models, hadronization of the parton system is
based on the Lund string-fragmentation scheme~\cite{lund}. 
The events of both types of samples
were passed through the OPAL detector simulation~\cite{simu}, processed
by the same reconstruction programs, and subjected to the same event
selection criteria as the data.

\subsubsection{Correction procedure}

To correct the data for the
finite acceptance and resolution of the detector, we determined
bin-by-bin correction factors for every bin of the distribution from
Monte Carlo samples with hadronization included, both before
(hadron level, ${H_i}^{\rm MC}$) and after (detector level,
${D_i}^{\rm MC}$) the detector simulation and selection cuts. The
hadron level treats all charged and neutral particles with lifetimes
greater than $3\times10^{-10}$ s as stable and has no initial state photon radiation. The correction factor for
detector effects for the $i^{th}$ bin of the distribution is given by:
\beq
{{C_i}^{\rm det}}=\frac{{H_i}^{\rm MC}}{{D_i}^{\rm MC}}\:.
\eeq

We examined the effects of initial state photon radiation by
switching the radiation on in the hadron level Monte Carlo simulation. The correction factors were found to be identical within statistical uncertainties with or without photon radiation over the entire phase space.

We estimated the hadronization effects by computing the distributions
of Monte Carlo events at the parton (${P_i}^{\rm MC}$) and hadron levels (${H_i}^{\rm MC}$) (before and after hadronization) and
defined the correction factor for hadronization by:
\beq
\label{cf_had}
{C_i}^{\rm had}=\frac{{P_i}^{\rm MC}}{{H_i}^{\rm MC}}\:.
\eeq

Using the two correction factors ${C_i}^{\rm det}$ and ${C_i}^{\rm had}$,
we correct the measured distribution to the parton level according to
\beq
{D_i}^{\rm corr}={C_i}^{\rm det}\,{C_i}^{\rm had}\,{D_i}^{\rm meas}\:.
\eeq
We show the distributions of the total correction factors,
${C_i}^{\rm tot}={C_i}^{\rm det}\,{C_i}^{\rm had}$, in Figs.~\ref{f:rates}
~and~\ref{f:angles}.

\subsection{Fit procedure}
\label{sec_fit}

Having prepared the corrected distribution of six variables, we
performed a $\chi^2$ minimization to determine the most probable values of the variables
$\eta=\as C_F/(2\pi)$, $x=C_A/C_F$ and $y=T_R/C_F$ (see
Section~\ref{app}) with the program MINUIT~\cite{minuit}. The function
$\chi^2$ was defined by:
\begin{center}
$\chi^2=\sum_{ij} \delta_i\,\sigma_{ij}^{-1}\,\delta_j$  ,  
\end{center}
where $\delta_i$ is the bin-wise difference between the theoretical QCD
prediction and the corrected data distribution, while $\sigma_{ij}$ is the
covariance matrix taking into account the statistical error of the
data, the finite Monte Carlo statistics, and the correlations
between the bins of the distributions.  The data and Monte Carlo event
samples were each divided into 90 subsamples\footnote{The number of subsamples is a result of optimization for the given data and number of bins.} to compute $\sigma_{ij}$,
using the standard formula:
\beq
\sigma_{ij}=\frac{1}{N\,(N-1)}\sum_{n=1}^N
\left(D_i^{(n)}-{\overline D}_i\right)\left(D_j^{(n)}-{\overline D}_j\right),
\eeq
where $D_i^{(n)}$ is the value of the corrected distribution in bin $i$
of the $n$th subsample, ${\overline D}_i$ is the value in that bin for
the whole sample and $N$ is the number of subsamples.  We included bins
in the fit only if the total correction factors were in the range
$0.9<C_i^{\rm tot}<1.1$, except for the four-jet rate, where
$0.85<{C_i}^{\rm tot}<1.15$ was used.\footnote{There are not enough
bins in the smaller range to obtain a reliable fit.} This bin choice is
motivated by our expectation that the Monte Carlo estimate is less
reliable if hadronization and/or detector distortions are large. The
total number of bins used in the fit was 82, leaving 79 degrees of
freedom since there are three fitted parameters ($\eta$, $x$ and $y$, see
Section~3.5).

\subsection{Fit results}

The central values of the fit results for the parameter $\eta$ and colour
factor ratios $x$ and $y$ are listed in the upper section of
Table~\ref{t:globalfit}.  The renormalization scale was fixed at
$x_{\mu}=1$.  The errors and correlations ($\rho$) in the table are
purely statistical.

We observe a strong correlation between $x$ and $y$.  We tested other
variables (for instance, the three-jet rate) in an attempt to decrease
these correlations, but did not observe a significant improvement. The
correlations were taken into account when the fit results were
converted to the standard QCD parameters $\alpha_s(M_Z)$ and colour
factors using the definitions in Eqs.~(\ref{2loopas}) and (\ref{xydef})
and $T_R=1/2$.  These latter results are given in the lower section of
Table~\ref{t:globalfit}.
\begin{table}[!htb]
  \begin{center}
    \begin{tabular}{|c||c|} \hline 
     $\eta=\as C_F/(2\pi)$      & 0.0256$\pm$0.0003  \\
     $x=C_A/C_F$         &   2.25$\pm$0.08    \\ 
     $y=T_R/C_F$         &   0.37$\pm$0.04    \\ 
$\rho_{\eta\,x}$ &  -0.33             \\
$\rho_{\eta\,y}$ &  -0.11             \\ 
$\rho_{x\,y}$    &   0.90             \\
$\chisqd$        &  98.5/79           \\
\hline \hline
 $\as$           & 0.120$\pm$0.011    \\
 $C_A$           & 3.02$\pm$0.25      \\
 $C_F$           & 1.34$\pm$0.13      \\
\hline 
    \end{tabular}
  \end{center}
  \caption[]{
Results of the default fit. The errors and correlations are statistical only.}
\label{t:globalfit}
\end{table}

\section{ Systematic uncertainties }
\label{sec_syserr}
The systematic uncertainties were evaluated by
considering the following effects:
\begin{itemize}
\itemsep -2pt
\item the measurement process and accuracy of the Monte Carlo detector
simulation,
\item dependence on the model of hadronization,
\item theoretical uncertainties arising from the choice of the renormalization
  scale,
\item variation of the matching scheme in the case of the
differential two-jet distribution,
\item variation of the fit range,
\item the background from five parton events,
\item variation of the parameter $y_{\rm cut}$.

\end{itemize}
In the systematic checks we always used the same covariance matrix
    that was optimized for the standard analysis.
\subsection{Detector effects}

The uncertainty related to the detector simulation and the
measurement process was estimated by repeating the analysis using
either tracks only or clusters only for the
data and the detector level Monte Carlo samples. For the tracks
and clusters we applied the same selection procedure as
described in Section~3.2, except that all cluster related criteria were left out in the former case and all track related criteria were left out
in the latter. Furthermore, the entire analysis was repeated using
$|\cos\theta_\thrust| < 0.7$ (rather than $|\cos\theta_\thrust|<0.9$),
or $N_{ch}\geq 8$ (rather than $N_{ch}\geq 5$).  The results obtained
for these analyses are given in Table~\ref{t:global_det}.
\begin{table}[!htb]
  \begin{center}
    \begin{tabular}{|l||cccc|} \hline 
     & $\eta$           & $x$ & $y$        & $\chisqd$     \\ \hline\hline
     tracks only& 0.0257$\pm$0.0003 & 2.24$\pm$0.08& 0.37$\pm$0.04  & 127.1/79   \\ 
     clusters only&       0.0256$\pm$0.0003 & 2.25$\pm$0.08& 0.38$\pm$0.04  & 124.2/79    \\
     $|\cos\theta_\thrust|<0.7$&  0.0256$\pm$0.0003 & 2.26$\pm$0.08& 0.38$\pm$0.04  & 117.9/79    \\ 
     $N_{ch}\geq 8$&         0.0256$\pm$0.0003 & 2.25$\pm$0.08& 0.37$\pm$0.04  & 98.3/79    \\ \hline \hline
     standard analysis & 0.0256$\pm$0.0003 & 2.25$\pm$0.08& 0.37$\pm$0.04  & 98.5/79    \\ \hline
    \end{tabular}
  \end{center}
  \caption[]{
Fit results obtained by varying systematic conditions related to detector effects. The errors are statistical only.}
\label{t:global_det}
\end{table}

\subsection{Hadronization models}

Several combinations of different Monte Carlo models
were considered to estimate the uncertainty associated
with the hadronization correction.  Since we used two different
Monte Carlo samples to determine the correction factors (ME and PS), we
examined the effects induced by variations of these models separately.

The angular correlations were corrected using the ME version of JETSET
with the Lund string fragmentation scheme, as stated in Section 3.3.1. In
the systematic studies we used the ME version of the HERWIG generator
\cite{herwig}, which produces four-parton final states according to the leading order QCD matrix elements, and then a parton shower is started from each of the four partons followed by hadronization based on the cluster fragmentation.\footnote{The HERWIG-ME has not been tuned to OPAL data. We used CLSMR(1)=0.4, PSPLT(2)=0.33, DECWT=0.7, and parameters not listed here were left at their default values.} We also varied the
principal hadronization parameters of the JETSET ME Monte Carlo
program, $\sigma_q$ and $a$. As the hadronization parameter $b$ is
strongly correlated with $a$ we kept it fixed.  We increased $a$ or
decreased $\sigma_q$, alternately, by 10\,\%.  These changes correspond
to about one standard deviation in the parameter values allowed for the
JETSET ME generator at $Z$ peak, as found by the LEP
experiments~\cite{MEpars_LEP}.  The fits were redone using data
corrected by the factors calculated from these Monte Carlo samples at
the generator level (reevaluating $C_i^{\rm had}$ only) and using the
standard $C_i^{\rm det}$ correction factors for detector effects.  

The effect of the hadronization process
on the four-jet rates and differential two-jet rates was studied by
using the PS version of HERWIG instead of the PS version of JETSET and by
varying the Lund hadronization parameters $\sigma_q$ and $a$ as for the
angular correlations. The parameter $Q_0$ defining the end of the
parton shower cascade was also varied from its standard value of
$1.9$ GeV to $2.4$ GeV. The theoretical predictions assume 
zero quark masses. To assess the effect of this assumption, we determined the
hadronization correction, $C^{\rm had}$, using JETSET Monte Carlo
events with only light primary quarks $(u, d, s, c)$ at the parton
level but with all available flavours $(u, d, s, c, b)$ at the hadron
level.  The results provided by these variations are presented in
Table~\ref{t:global_had}.
\begin{table}[!htb]
  \begin{center}
    \begin{tabular}{|l||cccc|} \hline 
     & $\eta$           & $x$ & $y$        & $\chisqd$     \\ \hline\hline
     HW-ME+HW-PS   & 0.0245$\pm$0.0003 & 2.43$\pm$0.18 & 0.44$\pm$0.08 & 274.5/79 \\    
     JT-ME($a+1\sigma$)+JT-PS$^*$&        0.0258$\pm$0.0003      & 2.15$\pm$0.08   & 0.33$\pm$0.04   & 94.5/79      \\ 
     JT-ME($\sigma_q-1\sigma$)+JT-PS$^*$& 0.0257$\pm$0.0003      & 2.16$\pm$0.08   & 0.33$\pm$0.03   & 156.5/79     \\
     JT-ME$^*$+JT-PS($a+1\sigma$)&        0.0253$\pm$0.0003      & 2.28$\pm$0.08   & 0.37$\pm$0.04   & 106.6/79  \\    
     JT-ME$^*$+JT-PS($\sigma_q-1\sigma$)& 0.0253$\pm$0.0003      & 2.28$\pm$0.08   & 0.37$\pm$0.04   & 106.1/79   \\
     JT-ME$^*$+JT-PS($Q_0+1\sigma$)&      0.0252$\pm$0.0003      & 2.29$\pm$0.08   & 0.37$\pm$0.04   & 101.1/79    \\
     JT-ME$^*$+JT-PS($udsc$)&         0.0260$\pm$0.0003  & 2.19$\pm$0.07   & 0.33$\pm$0.03   & 173.1/79     \\ \hline \hline
     standard analysis & 0.0256$\pm$0.0003 & 2.25$\pm$0.08& 0.37$\pm$0.04  & 98.5/79    \\ \hline
    \end{tabular}
  \end{center}
  \caption[]{Fit results obtained by varying the Monte Carlo models used
for the hadronization corrections.  The abbreviations are as follows:
$*$ denotes the Monte Carlo model used in the standard analysis, the model
before '+' is used to define hadronization corrections for the angular
correlations while the model after the '+' is used to define
hadronization corrections for the jet rates.  JT-ME: JETSET 7.4 matrix
element generator with string fragmentation, JT-PS: JETSET 7.4 parton
shower generator with string fragmentation, HW-ME: HERWIG 6.1 matrix
element generator with cluster fragmentation, HW-PS: HERWIG 5.9 parton
shower generator with cluster fragmentation.}
\label{t:global_had}
\end{table}

\vspace*{-0.05cm}
\subsection{Renormalization scale ambiguity}

The arbitrariness in the choice of the renormalization scale $\mu$
leads to an ambiguity in the theoretical prediction. We perform our default analysis at the scale factor \mbox{$x_{\mu}=\mu/E_{\rm cm}=1$.}

\begin{figure}[!htb]
\begin{center}
\hspace*{-7cm}
\epsfig{file=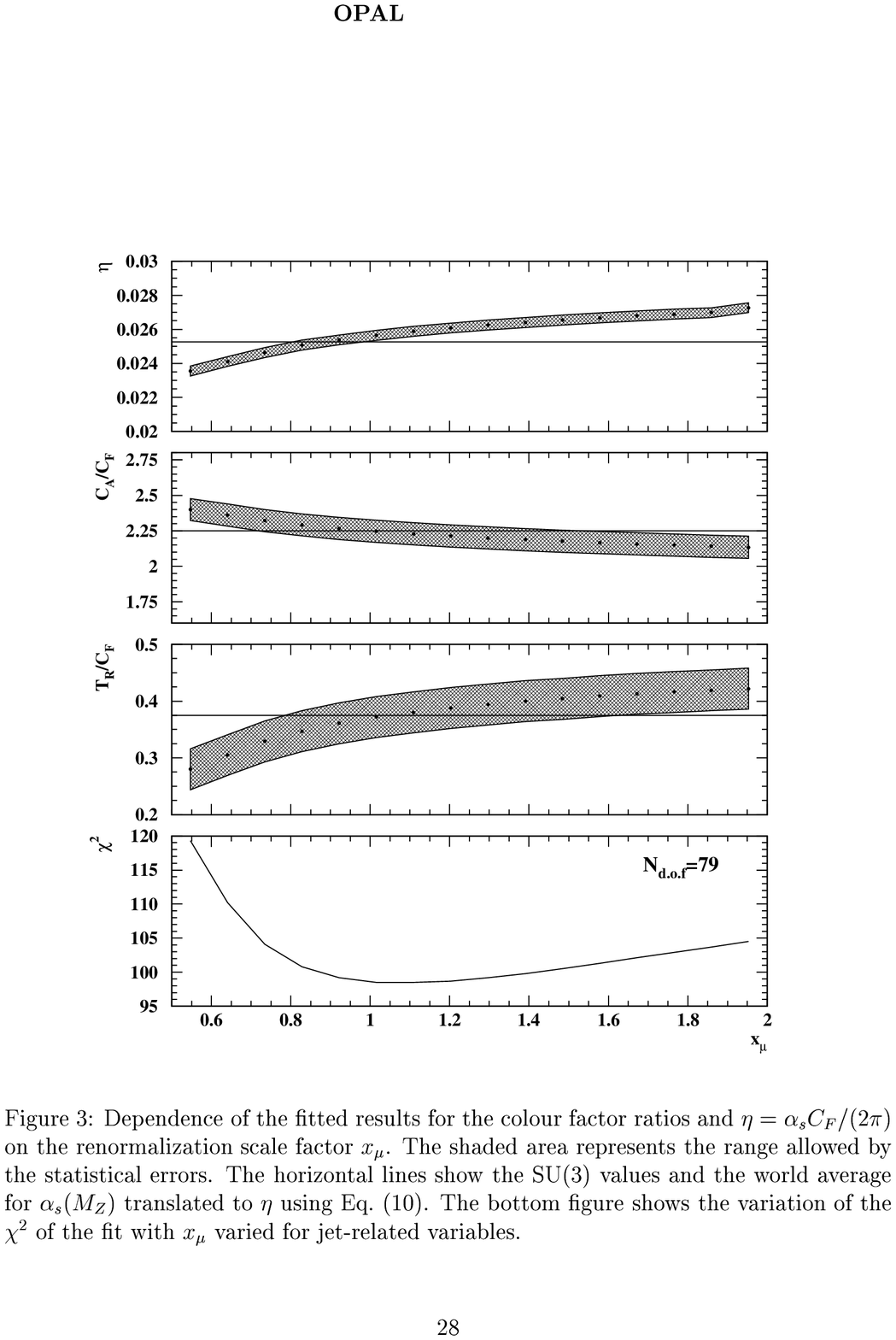,clip,width=15mm}\\
\vspace*{-2.2cm}
\resizebox{\textwidth}{!}
{\includegraphics{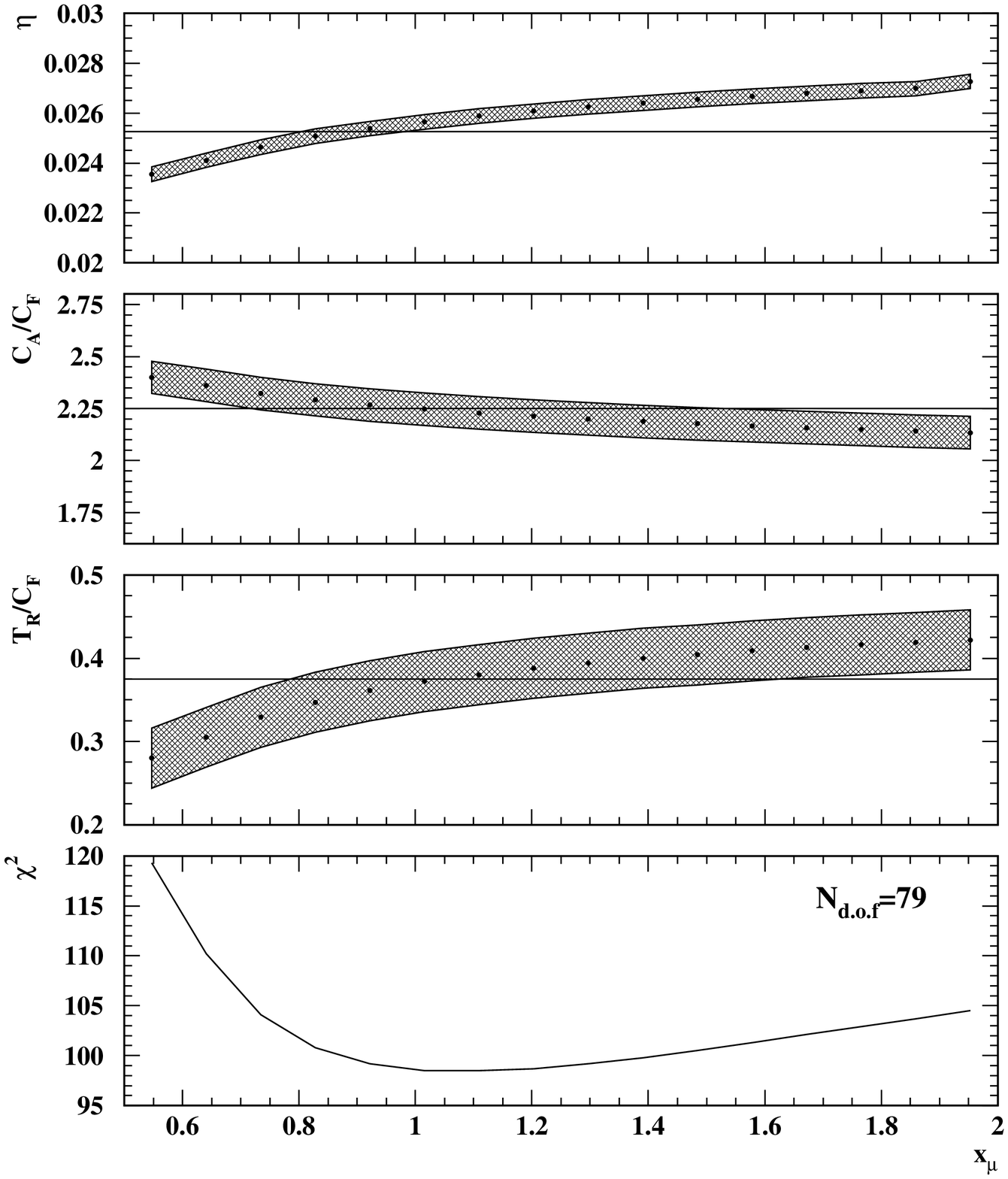}}
\caption[mugl]{Dependence of the fitted results for the colour factor ratios and $\eta=\as
C_F/(2\pi)$ on the renormalization scale factor $x_{\mu}$. 
The shaded area represents the range allowed by the statistical errors.
The horizontal lines show the SU(3) values and the world average for
$\as(M_Z)$ translated to $\eta$ using Eq.~(\ref{2loopas}). The bottom
figure shows the variation of the $\chi^2$ of the fit with $x_{\mu}$
varied for jet-related variables.}
\label{f:gl_xmu}
\end{center}
\end{figure}

To estimate the systematic error due to the scale ambiguity,
we varied the renormalization scale within the range $0.5\leq x_{\mu}\leq 2$. Varying $x_{\mu}$ probes the effect of missing higher orders, and these could be very different for the NLO and NLLA calculations.
We performed two variations separately: varying $x_\mu$ for the
jet-related variables while keeping it fixed at the default value
for the four-jet angular correlations, or vice versa.
Table~\ref{t:global_theorange} presents the results of this systematic
check.  As expected, the normalised angular correlations were not
sensitive to the scale choice, but the jet related variables showed
significant sensitivity.  In Fig.~\ref{f:gl_xmu} we show the dependence
of the $\chi^2$ of the fit on $x_{\mu}$, for the case of jet-related
variables with $x_{\mu}=1$ kept fixed for the angular correlations.
The minimum of the $\chi^2$ is seen to lie close to $x_{\mu}=1$.

\subsection{Matching ambiguity}

For the differential two-jet rates, we performed the fit using the 
$R$-matching scheme to combine the NLL and NLO approximations rather than the
$\ln R$-matching.
The result of this variation is listed in Table~\ref{t:global_theorange}.

\subsection{Fit range}

The uncertainty related to the number of bins selected for the fit was
evaluated by repeating the fit using those bins for which the values
of the correction factors were in the range of $0.8<{C_i}^{\rm
tot}<1.2$ instead of the default range $0.9<{C_i}^{\rm tot}<1.1$ or
$0.85<{C_i}^{\rm tot}<1.15$. The results are presented in
Table~\ref{t:global_theorange}.  The variation of the fit range for the
jet-related variables does not influence the value of the $\chisq$
significantly. The fit-range variation for the angular variables \BZ\
and \amin\ is the main source of the much increased value of the $\chisq$.

\subsection{Five-parton background}

Five-parton events which are reconstructed as four-jet events
at the parton level are taken into account by the NLO theory. We
examined the effect of those five-parton events which yield four jets at
the detector level but five at the parton level. 
To reduce this five-parton background, we applied a cut on $y_{45}$, the
value of the resolution variable which separates the four- and five-jet
configurations of any given event. We optimized the value of this cut
by studying five-parton events generated with the PENTAJET Monte Carlo program
\cite{5jet}.  This program generates a five-parton configuration using
the same value of the intrinsic $y_0$ as the JETSET-ME program, then
hadronizes the partons using JETSET string fragmentation
with the ERT-MC-3 parameters of Ref.~\cite{param}. The
events were passed through the OPAL detector simulation program.  Of
20\,000 five-parton events, 2\,517 were found to be four-jet events at
detector level, but five-jet events at parton level, for $\ycut =0.008$.
The number of these events reduces to 376 when we keep events with $y_{45}<0.003$, eliminating 85\,\% of the five-parton background, while leaving 60\,\% of the data.

The results of the fit with this additional requirement are
presented in Table~\ref{t:global_theorange}.

\subsection{Dependence on the parameter $y_{\rm cut}$}

The value of the jet resolution parameter \ycut, used in selecting
four-jet events to define the angular correlations, can be chosen
arbitrarily, as long as it is safely larger than the intrinsic $y_0$ of the
JETSET-ME program.  In order to have a high statistics sample, our
default choice was the smallest possible value, which allows us to
examine the stability of the results when \ycut\ is increased. The
four-jet event sample decreases very rapidly with increasing \ycut,
which constrains the region in \ycut\ where a reliable systematic check
can be performed.  We repeated the entire analysis using $\ycut = 0.01$.
The results of this fit are presented in Table~\ref{t:global_theorange}.

\begin{table}[!htb]
  \begin{center}
    \begin{tabular}{|l||cccc|} \hline 
     & $\eta$           & $x$ & $y$        & $\chisqd$     \\ \hline\hline
     $x_{\mu}=1.0$ for angular corr. and & & & & \\
     $x_{\mu}=0.5$ for $R_4$ and $D_2$  &                   0.0235$\pm$0.0003  & 2.40$\pm$0.08   & 0.28$\pm$0.03   & 118.0/79        \\ 
     $x_{\mu}=2.0$ for $R_4$ and $D_2$&                   0.0271$\pm$0.0003  & 2.14$\pm$0.08   & 0.42$\pm$0.04   & 103.7/79       \\ \hline 
     $x_{\mu}=1.0$ for $R_4$ and $D_2$ and& & & & \\
     $x_{\mu}=0.5$ for angular corr.  &                   0.0256$\pm$0.0003  & 2.25$\pm$0.08   & 0.37$\pm$0.03   & 98.2/79        \\ 
     $x_{\mu}=2.0$ for angular corr.&                   0.0257$\pm$0.0003  & 2.24$\pm$0.08   & 0.37$\pm$0.04   & 99.0/79       \\ \hline 
    R-matching for $D_2$     &        0.0281$\pm$0.0005  & 2.06$\pm$0.09   & 0.40$\pm$0.03   & 103.4/79 \\\hline \hline  
     $0.8<{C_i}^{\rm tot}<1.2$       &0.0255$\pm$0.0002  & 2.28$\pm$0.06   & 0.39$\pm$0.03   & 339.6/96 \\\hline \hline
     5-parton background&                                   0.0257$\pm$0.0003&2.18$\pm$0.08  & 0.34$\pm$0.04  &112.5/79 \\ \hline \hline
     $y_{\rm cut}=0.01$&0.0257$\pm$0.0003 & 2.27$\pm$0.21 &0.38$\pm$0.09
     &127.1/79 \\ \hline \hline
     standard analysis & 0.0256$\pm$0.0003 & 2.25$\pm$0.08& 0.37$\pm$0.04  & 98.5/79    \\ \hline
    \end{tabular}
  \end{center}
  \caption[]{Fit results obtained by varying the renormalization scale, the
     matching scheme for $D_2$, the fit range, the 5-parton background and the
     $y_{\rm cut}$ parameter.}
\label{t:global_theorange}
\end{table}

\subsection{Additional systematic checks}

We tested the influence of the two- and three-parton events that give
rise to four-jet events at the detector level by repeating the analysis
requiring $E_{\rm min} = 5\,$GeV on the minimum energy of the jets
instead of the default value $E_{\rm min}=3\,$GeV.  A higher value of
the jet energy reduces the number of two- and three-parton events which
satisfy the four-jet selection criteria in the detector. We generated 
1~million events with two, three and four partons in the initial state
using the ME version of JETSET with renormalization scale set to
$x_{\mu}^2=0.002$, which provides the correct proportion of two, three
and four jet events for $y_{\rm cut}=0.008$.  We then counted the
proportion of the two- and three-parton events which passed the
four-jet selection at detector level.  With $E > E_{\rm min} = 5\,$GeV,
there are no background events from the two- and three-parton events.
With $E > E_{\rm min}=3\,$GeV, a background of 1\,\% appears (519 two- or
three-parton events in the sample of 53\,302 four-jet events selected
at the detector level).  We found negligible effects on the final
results of the fit, therefore, we did not include the effect of this
check in the systematic errors. 

As stated in Section~2, the fits to individual angular correlations give
rather ambiguous results. Therefore, we choose to use all four angular
correlation variables in our standard analysis. As a consistency check,
we performed the fit using only five variables instead of six, removing
one of the six variables at a time.  The results of these fits are
presented in Table ~\ref{t:5var_fit}. The errors are statistical only.
We found that leaving out any one of the observables yields consistent
results for the measured parameters within the statistical errors of
the measurements. Therefore, we did not include the effect of this
check in the systematic errors.

The number of light quark flavours in a fixed order calculation is usually set to $N_f=5$ at LEP energies because the hard scattering scale is assumed to be the \epem\ center of mass energy. However, one may argue that in the four-quark channel the $g\rightarrow b\bar{b}$ splitting is suppressed compared to the $g\rightarrow q\bar{q}$ (q=u, d, s, c) splittings, because the scale of that process is much lower than the scale at the primary vertex. To study the effect of suppressed $g\rightarrow b\bar{b}$ splitting on our results, we also used $N_f=4$ light flavours at the secondary vertex in the fixed order prediction for the angular correlations. The results of this check are also shown in Table ~\ref{t:5var_fit}. 
The colour factor ratios increase slightly, remaining well within the statistical errors. The quality of the fit is almost unchanged. We did not include the effect of this check in the systematic errors, because it is not customary to do so.

\begin{table}[!htb]
  \begin{center}
    \begin{tabular}{|l||cccc|} \hline 
                    & $\eta$             & $x$
                    & $y$& $\chisqd$\\ \hline\hline
      without                  &                  &              &              &          \\\hline           
     Bengtsson-Zerwas angle    & 0.0257$\pm$0.0003& 2.22$\pm$0.17&0.36$\pm$0.08 & 63.7/64 \\\hline
     Nachtmann-Reiter angle    & 0.0256$\pm$0.0003& 2.39$\pm$0.20&0.44$\pm$0.09 & 56.0/62 \\ \hline
     K\"orner-Schierholtz-Willrodt angle& 0.0256$\pm$0.0003&2.29$\pm$0.21 &0.39$\pm$0.09 & 63.1/63  \\ \hline
     angle of the two softest jets&0.0256$\pm$0.0003&2.31$\pm$0.17&0.40$\pm$0.08& 57.9/65  \\ \hline 
     four-jet rates            & 0.0243$\pm$0.0007&2.18$\pm$0.09&0.39$\pm$0.04& 75.8/73 \\ \hline
     differential two-jet rates& 0.0251$\pm$0.0010&2.31$\pm$0.12 & 0.36$\pm$0.04& 52.1/65\\  \hline \hline
     $N_f=4$                   & 0.0256$\pm$0.0003& 2.28$\pm$0.08&0.39$\pm$0.04&95.9/79\\\hline\hline
     standard analysis & 0.0256$\pm$0.0003 & 2.25$\pm$0.08& 0.37$\pm$0.04  & 98.5/79    \\ \hline
    \end{tabular}
  \end{center}
  \caption[]{Fit results using five variables only or using $N_F=4$.}
  \label{t:5var_fit}
\end{table}

\subsection{Total systematic error}
To combine the effects of the systematic variations we employed a Bayesian
method that was developed in a similar analysis by the ALEPH collaboration
~\cite{color_ALEPH2}, where the explicit formulae we used can be found.
The basic idea is that the information obtained from a fit is accepted or
rejected based on the quality of the fit, i.e. the magnitude of the $\chi
^2$. From a Bayesian point of view we presume that all models are equally
well suited to the analysis. However, a large $\chisqd$ indicates that the
probability of a model is low and, therefore, it should contribute to the
systematic errors with a small weight.  Accordingly, the systematic error
corresponds to an increase of the $\chi^2$ by 1. 

We could not use the
Bayesian method in two cases. In the variation of the fit range the number
of bins changed, while in checking the renormalization scale uncertainty
for the angular correlations we did not find a pronounced minimum of the
$\chisq$ curve. Furthermore we did not use the Bayesian method to estimate the systematic
uncertainty from the five-parton background because the event
statistics for this check were modified significantly, by about 40\%,
relative to the standard analysis. 
In these cases we estimated the systematic error by the
deviation of the results from the standard values.

The errors
obtained from the different systematic checks are summarized in
Table~\ref{t:global_totsys}. The theoretical uncertainty refers to the
larger of the errors from the scale uncertainty of the jet-related
variables, the matching ambiguity
(both of those estimate the effect of
unknown higher order perturbative contributions), and the
scale uncertainty of the four-jet angular variables. The different
sources were added in quadrature to define the total systematic
error.\footnote{If the Bayesian approach were not adopted and the
systematic errors were estimated by the largest deviation from the
standard values, then we obtain the following systematic errors:
$\Delta \eta=0.0027,\,\Delta x=0.27,\,\Delta y=0.08$. For the basic
QCD parameters these errors are: $\Delta \as=0.030,\,\Delta
C_A=0.72,\,\Delta C_F=0.31$.}

Taking into account the systematic correlations presented in
Table~\ref{t:global_totsys}, we converted these values to the errors of the
standard QCD parameters $\alpha_s(M_Z)$ and colour factors using the
definitions in Eq.~(\ref{2loopas}) and (\ref{xydef}) and $T_R=1/2$. 
The result of the conversion is presented in Table~\ref{t:global_totsys}.
The errors are dominated by the theoretical uncertainty.
\begin{table}[!htb]
  \begin{center}
    \begin{tabular}{|l||cccccc|} \hline 
                    & $\Delta\eta$             & $\Delta x$
                    & $\Delta y$&$\rho_{\eta x}$&$\rho_{\eta y}$ &$\rho_{ xy}$\\ \hline\hline
     Detector effects          & 0.0000       & 0.004 & 0.002 & -0.87 & -0.89 & 0.95 \\\hline
     Hadronization             & 0.0003       & 0.056 & 0.021 & -0.86 & -0.71 & 0.92\\ \hline
     Theoretical uncertainty   & 0.0013       & 0.096 & 0.038 & -1.0  & 1.0  & -1.0 \\ \hline
     Fit range                 & 0.0001       & 0.033 & 0.013 & -1.0  & -1.0  & 1.0 \\ \hline
     5-parton background          & 0.0001       & 0.070 & 0.039 & -1.0  &-1.0  & 1.0 \\ \hline
     $y_{\rm cut}$             &0.0000  & 0.005 & 0.002 &1.0 &1.0 &1.0 \\ \hline \hline
     Total systematic uncertainties&0.0013    & 0.136 & 0.060 & -0.83 & 0.06  & 0.38  \\\hline \hline
                    & $\Delta \as$             & $\Delta C_A$
                    & $\Delta C_F$&&&\\ \hline\hline
     Total systematic uncertainties&0.020& 0.49 & 0.22 & & & \\  \hline
    \end{tabular}
  \end{center}
  \caption[]{Contributions to the total systematic error.}
  \label{t:global_totsys}
\end{table}

\section{Results and Discussion}

\begin{figure}[!htb]
\begin{center}
\hspace*{-1cm}
\epsfig{file=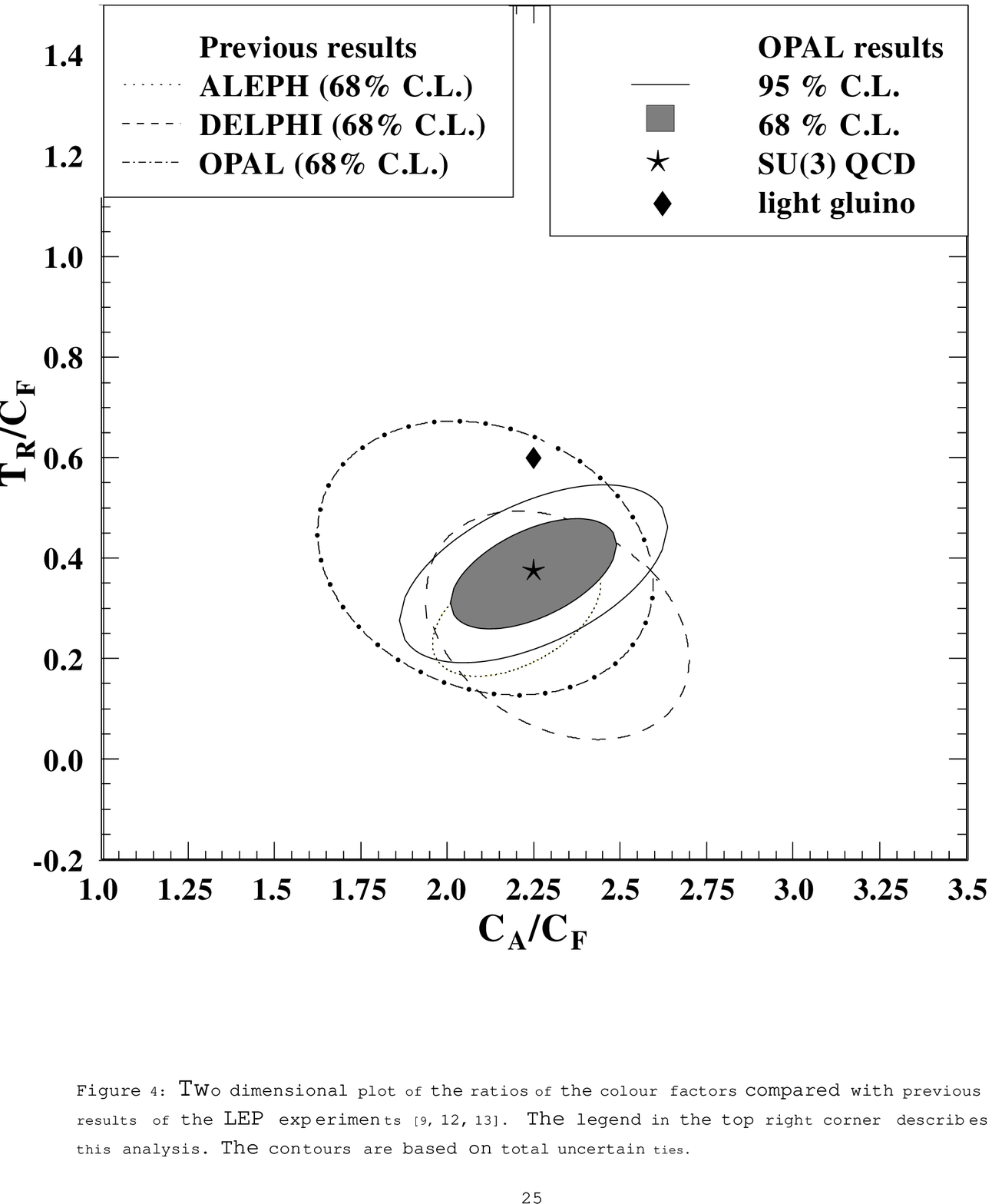,clip,width=173mm}
\caption[mugl]{Two dimensional plot of the ratios of the colour factors
compared with previous results of the LEP experiments
\cite{color_DELPHI,color_ALEPH2,color_OPAL}. The legend in the
top right corner describes this analysis. The contours are based on total uncertainties. } \label{f:final}
\end{center}
\end{figure}

The procedure outlined above yields 

\[
C_A/C_F=2.25 \pm  0.08\stat \pm 0.14\syst\:,\quad
T_R/C_F=0.37 \pm  0.04\stat \pm 0.06\syst \:.
\]
Repeating the entire analysis with $\as(M_Z)$ fixed to its world
average value 0.1184~\cite{as} leads to similar results:
\[
C_A/C_F=2.29 \pm  0.06\stat \pm 0.14\syst\:,\quad
T_R/C_F=0.38 \pm  0.03\stat \pm 0.06\syst \:.
\]
To compare the results to previous measurements, we
show in Fig.~\ref{f:final} the two dimensional 68\% and 95\% confidence level contour plots of the
colour factor ratios based on total uncertainties obtained by our
analysis. 

The DELPHI Collaboration~\cite{color_DELPHI} has performed a least-squares fit of leading order predictions to the two-dimensional distribution in the variables \NR\ and \amin\ in order to measure of the colour factor ratios. The DELPHI Collaboration~\cite{color_DELPHI2} also performed this analysis with b-tagging to separate quark jets from gluon jets which increases sensitivity.
The OPAL Collaboration fitted the LO prediction to a three-dimensional distribution of the \NR\ ,  \amin\ and \BZ\ variables~\cite{color_OPAL}. In a second measurement the OPAL Collaboration used event shape distributions of LEP1 data for fitting \as\ and one of the colour factors~\cite{shapes_OPAL}.
Furthermore, another recent measurement of the QCD colour factors and \as\ using event shape distributions at several energy points and power corrections was carried out~\cite{shapes_OPAL2}. The ALEPH Collaboration~\cite{color_ALEPH2} was the first to present a simultaneous measurement of the strong coupling and the colour factors. The distribution $D_2$ and all four angular distributions defined in Section~2 have been employed.  
The results of the ALEPH
experiment were determined using leading order predictions for the
four-jet observables.
The systematic uncertainties include contributions from variation of the renormalization scale, the matching ambiguity, from hadronization and detector effects, and finally from the estimation of mass effects. 
Our results are in agreement with
these previous results.
The main new
feature of our analysis is that we obtained our results using NLO predictions for four-jet angular correlations. Furthermore, our
systematic checks contain additional items arising from the estimate of the
effects of the five-parton background and the variation of the
$y_{\rm cut}$ parameter. We also checked that assuming four rather than five 
light flavours at the secondary vertex does not influence our results. This effect was not examined in any of the previous studies. 

With the inclusion of the higher order
corrections, the colour factor ratios became better constrained,
especially $T_R/C_F$. We found an increase of about 15\% on $T_R/C_F$ using the NLO theoretical predictions instead of the LO one, while $C_A/C_F$ was only slightly affected (about 3\%). 

The diamond symbol in Fig.~\ref{f:final} indicates the result one would expect if a
leading order perturbative theory with light gluinos was the
correct theory. This point falls well outside our 95\% C.L.
contour, indicating that the data do not favour the existence of light
gluinos. A complete analysis to test for the existence of gluinos
should use hadronization models and NLO predictions that include the
effect of the gluinos~\cite{gluino}. The latter is possible using the
{\tt DEBRECEN} partonic Monte Carlo program we used for generating the
NLO results (see the Appendix), but presently none of the
existing hadronization models include gluinos.

Finally, we converted our measured parameters to the standard QCD
parameters (using $T_R=1/2$), which leads us to our main results:
\begin{eqnarray*}
&&
\label{final}
\as=0.120 \pm  0.011\stat \pm 0.020\syst \\ &&
C_A=3.02 \pm  0.25\stat \pm 0.49\syst \\ &&
C_F=1.34 \pm  0.13\stat \pm 0.22\syst 
\end{eqnarray*}

Other techniques~\cite{as} provide precise determinations of 
\as\ . In this study the measurement of \as\ serves as a crosscheck and a consistent value was found.
The larger uncertainty we obtain for \as\ compared to other techniques can be traced to the functional dependence on the other two parameters.

\vspace*{-0.5cm}
\section{Summary}

In this paper, a test of perturbative QCD at LEP has been presented. 
With only the assumption of non-abelian gauge symmetry and standard
hadronization models, we measured the eigenvalues of quadratic Casimir
operators --- the colour factors --- of the underlying gauge group,
together with the coupling of QCD. The measurement was based on
the comparison of next-to-leading order perturbative predictions
for four-jet angular variables, and the resummation improved
next-to-leading order perturbative predictions for the multi-jet rates,
to OPAL data taken between 1991 and 1995 at the \znull\ peak. The
results for the colour factors are as follows:
\[
C_A=3.02 \pm  0.25\stat \pm 0.49\syst\:,\quad
C_F=1.34 \pm  0.13\stat \pm 0.22\syst\:,
\]
from a global fit to the angular correlations, four-jet rates, and
differential two-jet rates. These values are in agreement with SU(3)
values of $C_A = 3$ and $C_F = 4/3 $ as well as with measured
values obtained previously at LEP. The corresponding strong coupling
value,
\[
\as=0.120 \pm  0.011\stat \pm 0.020\syst\:,
\]
is in agreement with the world average.  
\appendix
\section{Appendix: Theoretical Predictions}
\label{app}
The NLO perturbative expansion of the differential two-jet rates has the
following expression:
\beq\label{D2}
D_2(y_{23})\equiv
\frac{1}{\sigma_{\rm tot}}\,\frac{\d \sigma}{\d y_{23}}
=\eta(\mu)A'_{D_2}(y_{23})
+\eta(\mu)^2\left[B'_{D_2}(y_{23})
+\beta_0\,\ln(x_\mu)A'_{D_2}(y_{23})\right]\:,
\eeq
where $y_{23}$ is the \ycut\ value which separates the two- and
three-jet configurations of an event. The corresponding formula in the
case of the four-jet rates takes a similar form:
\beq\label{R4}
R_4(\ycut)\equiv
\frac{\sigma_{4-{\rm jet}}}{\sigma_{\rm tot}}
=\eta(\mu)^2\,B_{R_4}(\ycut)
+\eta(\mu)^3\left[C_{R_4}(\ycut)
+\beta_0\,\ln(x_\mu^2)B_{R_4}(\ycut)\right]\:.
\eeq
In Eqs.~(\ref{D2}) and (\ref{R4}) $\sigma_{\rm tot}$ is the total
hadronic cross section at O($\eta$) accuracy and
\beq
\label{2loopas}
\eta(\mu)\equiv \frac{\asmu\,C_F}{2\pi}
=\frac{\eta(M_Z)}{w(\mu,M_Z)}
\left(1-\frac{\beta_1}{\beta_0}\,\eta(M_Z)
\,\frac{\ln(w(\mu,M_Z))}{w(\mu,M_Z)}\right)\:.
\eeq
In the second equality in Eq.~(\ref{2loopas}), we used the two-loop
expression for the running coupling with
\beq
w(\mu,M_Z)=1-\beta_0\,\eta(\mu)\ln\left(\frac{M_Z}{\mu}\right)\:,
\eeq
and $\mu$ denotes the renormalization scale ($x_{\mu}=\mu/\rs$, where
\rs\ is the total c.m. energy). We define the colour factor ratios as
\beq\label{xydef}
x=\frac{C_A}{C_F}\,,\quad y=\frac{T_R}{C_F}\,,\quad
y_f = N_f\,y\:,
\eeq
where $N_f$ is the number of active flavours. Then the coefficients in
the expansion of the $\beta$ function can be written as
\beq
\beta_0=\frac{11}{3}x-\frac{4}{3}y_f\:,\quad
\beta_1=\frac{17}{3}x^2-2\,y_f-\frac{10}{3}x\,y_f\:.
\eeq
The functions $A'_{D_2}$, $B_{R_4}$ and $B'_{D_2}$, $C_{R_4}$ in
Eqs.~(\ref{D2}) and (\ref{R4}) are the perturbatively
calculable coefficient functions in the Born approximation and the
radiative correction, respectively, which are independent of the
renormalization scale. The $A'_{D_2}$ function is independent of $x$
and $y$. The $B$ functions are linear forms and the
$C_4$ function is a quadratic form of these colour charge
ratios~\cite{NT4jets}:
\beq\label{Bn}
B_O=B^{(O)}_0+B^{(O)}_x\,x+B^{(O)}_y\,y\:,\qquad O = D_2\:,\;{\rm or}\; R_4
\eeq
\beq\label{Cn}
C_{R_4}=C^{(4)}_0+C^{(4)}_x\,x+C^{(4)}_y\,y+C^{(4)}_z\,z
+C^{(4)}_{xx}\,x^2+C^{(4)}_{xy}\,x\,y+C^{(4)}_{yy}\,y^2\:,
\eeq
where
\beq
z = \frac{C_C}{N_c C_F^3}\:,
\eeq
with
\beq
C_C = \sum_{a,b,c=1}^{N_c^2-1} \Tr(T^a T^b T^{\dag c})\,\Tr(T^{\dag c} T^b T^a)
\eeq
being a cubic Casimir invariant of the gauge group.\footnote{The results
of the Monte Carlo integration for $C^{(4)}_z$ are compatible with zero~\cite{NTqcd98},
therefore, $z$ is not included as a free parameter in our fits.}
We calculated these coefficient functions of the differential two-jet
rates in the $-6 \le \ln(y_{23}) \le 0$ $y_{23}$ range and those of
the four-jet rates in the $-3 \le \log_{10}(\ycut) \le 0$ \ycut\ range
using the partonic Monte Carlo program {\tt DEBRECEN}~\cite{debrecen}.
All theoretical predictions were obtained for $N_f=5$ light-quark
flavours and with the normalization $T_R=\frac{1}{2}$. The latter
choice is arbitrary and influences the values of the coefficient
functions in such a way that the physical prediction remains unchanged.

The resummation formulas for various event shapes are given in
Ref.~\cite{CTTWresum} at the NLL accuracy for the cumulative cross
section,
\beq
R(y) = \int_0^y\!\d y_{23}\,
\frac{1}{\sigma}\,\frac{\d \sigma}{\d y_{23}}\:,
\eeq
in the following form:
\beq
\label{nllad2}
R^{NLL}(L)=(1+C_1\eta+C_2\eta ^2)e^{L\,g_1(\eta L)+g_2(\eta L)}\:,
\eeq
where $L = -\ln(y_{23})$.
The functions $Lg_1(\eta L)$ and $g_2(\eta L)$ represent the sums of
the leading and next-to-leading logarithms, respectively, to all orders
in the strong coupling. The colour decomposition of the $g_1$ and $g_2$
functions is given in Ref.~\cite{DSansatz}. The coefficient $C_1 = -5/2 +
\pi^2/6 - 6\ln 2$ is known exactly \cite{CDFWresum} and $C_2$ can be
determined from numerical integration of the $\oaa$ matrix elements
(see below).

At NLO (i.e. $\oaa$ accuracy), the cumulative cross section for the
differential two-jet rates has the form:
\beq 
\label{fixd2}
R^{\oaa}(L)=1+A(L)\,\eta+B(L)\,\eta ^2\:.
\eeq
In order to have the best possible theoretical description, we fit the
experimental data to the matched NLL and NLO results, which is expected
to describe the data in a wider kinematical range than any of the two
separately. In the case of $D_2$, our default matching procedure is the
$\ln R$-matching, which combines the NLL and NLO approximations
according to the following formula:
\beeq
\label{lnRmatch}
&&
\ln R(L)=\Sigma(L ,\eta)
-(G_{11}\,L+G_{12}\,L^2)\eta-(G_{22}\,L^2+G_{23}\,L^3)\eta^2
\\ \nn && \qquad\qquad
+A(L)\,\eta
+\left(B_0(L)+x\,B_x(L)+y\,B_y(L)-\frac{1}{2}A(L)^2\right)\eta^2\:.
\eeeq
where $\Sigma(L , \eta)=Lg_1(\eta L)+g_2(\eta L)$ and the following two
terms represent the two lowest order terms in the $\eta$-expansion of
$\ln R$. In the actual analysis, we used the differential distribution
$\d\sigma/\d L$ that is obtained from Eq.~(\ref{lnRmatch}) as
\beeq
&&
\label{difflogR}
\frac{1}{\sigma}\frac{\d\sigma}{\d L}
=\exp(\ln R)\Bigg[\frac{\d\Sigma(L ,\eta)}{\d L}
-(G_{11}+2\,G_{12}\,L)\eta-(2\,G_{22}\,L+3\,G_{23}\,L^2)\eta^2
\\ \nn && \qquad\qquad
+A'(L)\,\eta
+\left(B_0'(L)+x\,B_x'(L)+y\,B_y'(L)-A(L)\,A'(L)\right)
\eta^2\Bigg]\:,
\eeeq
with $A'(L)$ and $B'(L)$  given in Eq.~(\ref{D2}).

Eq.~(\ref{difflogR}) contains the cumulative coefficients $A(L)$ and
$B_i(L)$, the latter ones implicitly in $R$, that cannot be directly
calculated using a partonic Monte Carlo program. We know, however, that
\beq 
\label{AL}
A(L)=C_1+\sum_{m=1}^2 R_{1m}\,L^m+D_1(L)\:,
\eeq
and
\beq
\label{BiL}
B_i(L)=C_2^{(i)}+\sum_{m=1}^4 R_{2m}^{(i)}\,L^m+D_2^{(i)}(L)\:,\qquad
i=0,x,y\:,
\eeq
where the colour decomposition of the $R_{2m}^{(i)}$ logarithmic
coefficients can be found in Ref.~\cite{as_OPAL_color}, except
for $R_{21}^{(i)}$.
This latter coefficient as well as the $C_2^{(i)}$ constants and the
$D_1(L)$, $D_2^{(i)}(L)$ functions can be determined by differentiating
Eqs.~(\ref{AL}) and (\ref{BiL}) with respect to $L$ and fitting to the
coefficient functions as obtained from the partonic Monte Carlo program
{\tt DEBRECEN}. Following the fitting procedure of
Refs.~\cite{CTTWresum,DSansatz}
we obtained $R_{21}^{(0)}=-16.531\pm 2.97\stat$,
$R_{21}^{(x)}=-4.293\pm 1.03\stat$, $R_{21}^{(y)}=4.54\pm 1.05\stat$,
$C_2^0=-25.56\pm 4.04\stat$, $C_2^x=-1.53\pm 1.12\stat$,
$C_2^y=-3.565\pm 1.17\stat $.

In order to account for the dependence on the renormalization scale
$x_{\mu}$, one has to make the following substitutions in
Eq.~(\ref{difflogR}):
\beeq
B(L)\to B(L)+A(L)\,\beta_0 \ln x_{\mu}\:,\\
\Sigma(\lambda, \eta)\to
\Sigma(\lambda, \eta)+\lambda^2 g_1^{'} \ln x_{\mu}\:,\\
G_{22}\to  G_{22}+G_{12} \beta _0\ln x_{\mu}\:.
\eeeq

The other widely used matching scheme is $R$-matching, which will
serve as a systematic check of our matching procedure. In this case
the NLL and NLO predictions are combined according to the following
formula~\cite{as_ALEPH,CDOTW}:
\beq\label{Rmatch}
R^{\rm R-match}=R^{\rm NLL}
+\left[\eta^{2}\left(B-B^{\rm NLL}\right)
      +\eta^{3}\left(C-C^{\rm NLL}\right)\right],
\eeq
where $B^{\rm NLL}$ and $C^{\rm NLL}$ are the coefficients in the
$\eta$-expansion of $R^{\rm NLL}$. In the case of multi-jet rates, the
NLL approximation does not exponentiate, therefore, the viable matching
scheme for the four-jet rates is $R$-matching. The resummation
formulas for the four-jet rates can be found in Refs.~\cite{CDOTW,NTqcd98}.

The general form of the NLO differential cross section for the four-jet
angular correlations $O_4$ is given by \cite{NTangulars}:
\beq\label{O4}
\frac{1}{\sigma_0} \frac{d\sigma }{\d O_4}(O_4)=
 \eta(\mu)^2\,B_{O_4}(O_4)
+\eta(\mu)^3
\left[C_{O_4}(O_4) + B_{O_4}(O_4)\,\beta_0\,\ln(x_\mu^2) \right]\:,
\eeq
where $\sigma_0$ denotes the Born cross section for the process
$e^+e^-\to \bar{q} q$. In Eq.~(\ref{O4}) the Born and correction functions
$B_{O_4}$ and $C_{O_4}$ have an analogous decomposition to
Eqs.~(\ref{Bn}) and (\ref{Cn}). To obtain the distributions
normalised to unity in perturbation theory, we calculated the coefficient
functions $B_i^{(O_4)}$ and $C_i^{(O_4)}$ in the linear and quadratic
forms using the partonic Monte Carlo program {\tt DEBRECEN}, and
normalised those with the total O($\eta^3$) four-jet cross section of
the same \ycut.


\def\ibid#1#2#3  {{\it ibid} {\bf #1}, #2 (19#3)}
\def\npb#1#2#3  {Nucl.\ Phys.\ B {\bf #1}, #2 (19#3)}
\def\npproc#1#2#3  {Nucl.\ Phys.\ B (Proc.\ Suppl.) {\bf #1}, #2 (19#3)}
\def\plb#1#2#3  {Phys.\ Lett.\ B {\bf #1}, #2 (19#3)}
\def\pl#1#2#3  {Phys.\ Lett.\ {\bf #1}, #2 (19#3)}
\def\prep#1#2#3  {Phys.\ Rep.\ {\bf #1}, #2 (19#3)}
\def\prd#1#2#3 {Phys.\ Rev.\ D {\bf #1}, #2 (19#3)}
\def\prl#1#2#3 {Phys.\ Rev.\ Lett.\ {\bf #1}, #2 (19#3)}
\def\zpc#1#2#3  {Zeit.\ Phys.\ C {\bf #1}, #2 (19#3)}
\def\cpc#1#2#3  {Comp.\ Phys.\ Comm.\ {\bf #1}, #2 (19#3)}
\def\ncim#1#2#3 {Nuovo Cim.\ {\bf #1}, #2 (19#3)}
\def\nciml#1#2#3 {Nuovo Cim.\ Lett.\ {\bf #1}, #2 (19#3)}
\def\jhep#1#2#3 {JHEP\ {\bf #1}, #2 (19#3)}
\def\nim#1#2#3 {Nucl.\ Instr.\ Meth.\ A {\bf #1}, #2 (19#3)}
\def\etal{{\em et al.}}

\newpage

\end{document}